\documentclass[final,1p,times]{elsarticle}
\usepackage{graphicx}
\usepackage{dcolumn}
\usepackage{bm}
\usepackage{xcolor}
\usepackage{subcaption}
\usepackage{amsmath}
\usepackage{hyperref}
\usepackage{natbib} \setcitestyle{comma,sort&compress,numbers,square}
\hypersetup{hypertex=true,
            colorlinks=true,
            linkcolor=blue,
            anchorcolor=blue,
            citecolor=blue}
\usepackage{amsthm}
\usepackage{geometry}
\usepackage{algpseudocode}
\usepackage{algorithm}
\theoremstyle{definition}

\newtheorem{definition}{Definition}

\DeclareMathOperator*{\argmax}{arg\,max}

\usepackage{xcolor}

\begin{document}
\begin{frontmatter}
\title{$\mathcal{H}$-HIGNN: A Scalable Graph Neural Network Framework with Hierarchical Matrix Acceleration for Simulation of Large-Scale Particulate Suspensions}
\date{}
\author[WISC]{Zhan Ma}
\author[WISC]{Zisheng Ye}
\author[WISC]{Ebrahim Safdarian}
\author[WISC]{Wenxiao Pan\corref{cor}}
\ead{wpan9@wisc.edu}
\cortext[cor]{Corresponding author}
\address[WISC]{Department of Mechanical Engineering, University of Wisconsin-Madison, Madison, WI 53706, USA}

\begin{abstract}
We present a fast and scalable framework, leveraging graph neural networks (GNNs) and hierarchical matrix ($\mathcal{H}$-matrix) techniques, for simulating large-scale particulate suspensions, which have broader impacts across science and engineering. The framework draws on the Hydrodynamic Interaction Graph Neural Network (HIGNN) that employs GNNs to model the mobility tensor governing particle motion under hydrodynamic interactions (HIs) and external forces. HIGNN offers several advantages: it effectively captures both short- and long-range HIs and their many-body nature; it realizes a substantial speedup over traditional methodologies, by requiring only a forward pass through its neural networks at each time step; it provides explainability beyond black-box neural network models, through direct correspondence between graph connectivity and physical interactions; and it demonstrates transferability across different systems, irrespective of particles' number, concentration, configuration, or external forces. While HIGNN provides significant speedup, the quadratic scaling of its overall \textit{prediction} cost (with respect to the total number of particles), due to intrinsically slow-decaying two-body HIs, limits its \textit{scalability}. To achieve superior efficiency across all scales, in the present work we integrate $\mathcal{H}$-matrix techniques into HIGNN, reducing the prediction cost scaling to quasi-linear. This $\mathcal{H}$-matrix accelerated HIGNN, termed $\mathcal{H}$-HIGNN, is built upon the following key components: 1) cluster-tree block partitioning using geometric information embedded in the graph; 2) cluster distance-to-size ratios for admissibility determination; and 3) adaptive cross approximation for low-rank approximations of admissible blocks. Through comprehensive evaluations, we validate $\mathcal{H}$-HIGNN's accuracy, and demonstrate its quasi-linear scalability and superior computational efficiency. It requires only minimal computing resources; for example, a single mid-range GPU is sufficient for a system containing 10 million particles. Finally, we demonstrate $\mathcal{H}$-HIGNN's ability to efficiently simulate practically relevant large-scale suspensions of both particles and flexible filaments. 
\end{abstract}

\begin{keyword}
Machine learning; Graph neural network; Hierarchical matrix; Particulate suspension; Hydrodynamic interaction; Flexible filament

\end{keyword}

\end{frontmatter}

\section{Introduction}

Particulate suspensions can play a key role in a wide range of scientific and engineering applications.
For example, studying the behaviors of suspension drops, involving particle clouds in viscous fluids, is important for a variety of industrial applications \cite{ParticleHI_PRL2019} and natural phenomena \cite{Manga1993Geophysical1,BUSH_THURBER_BLANCHETTE_2003}. If particles are linked by elastic bonding and bending forces as chains, one can model flexible filaments to better understand their deformations and interactions with viscous fluids in biological systems like DNA in flow, microorganismal swimming, and eukaryotic nuclear positioning \cite{schroeder2003observation,shelley2016dynamics,altunkeyik2025dynamics,du2019dynamics}. When dynamically actuated (e.g., by magnetic fields), particles or flexible filaments can be self-propelled and organized, thereby facilitating the design of micro-robots, soft robots, and micro-swimmers with potential applications in biomedicine, microfluidics, and biomimetic systems \cite{altunkeyik2025dynamics,manna2017colloidal,jin2021collective}. Therefore, understanding how particles or flexible filaments interact collectively and with viscous fluids, and how these interactions dictate macroscopic behaviors, is a subject of broad and significant interest. Accurate predictions of dynamic properties and behaviors of these systems are crucial for advancing this understanding. Besides external forces, the dynamics of particles is significantly impacted by hydrodynamic interactions (HIs) \cite{ColloidNS_NCM2019}. The incorporation of HIs poses challenges for simulations due to their influences across a wide range of spatial scales, spanning both short and long ranges, and the inherent many-body nature of HI, which renders pairwise interaction models inadequate \cite{Brady1988_SDreview,ParticulateReview_Maxey2017}. In low Reynolds number regimes, the dynamics of particles is primarily governed by viscous effects due to the significantly shorter inertial relaxation time compared to the viscous relaxation time. 

For simulating particulate suspensions at low Reynolds numbers, a wide array of numerical methods have been developed. A class of these methods relies on numerical discretizations and linear system solvers to handle the PDEs governing the bidirectional hydrodynamic couplings between the Stokes flow and particle kinematics,
such as 
immersed boundary method \cite{fogelson1988fast,IBM_CMAME2007}, boundary integral method \cite{ParticulateBI_JCP2020}, force-coupling method \cite{YEO2010FCM3}, and generalized moving least squares (GMLS) method \cite{GMLS_StokesColloid_HuCMAME2019,GMLS_ScalableSolver_Ye2022}, to name a few. Their high spatial and temporal resolution requirements, necessary for accurately capturing both long- and short-range HIs, limit their applicability to large-scale systems. For instance, GMLS's iterative adaptive refinement to resolve the fluid field between the near-to-contact particles significantly increases computational cost \cite{GMLS_StokesColloid_HuCMAME2019,GMLS_ScalableSolver_Ye2022}. Alternatively, the methods of Stokesian Dynamics (SD) \cite{Brady1988_SDreview}, reduce computational burden by directly constructing the mobility tensor (matrix), bypassing the need to solve PDEs. This mobility tensor, unlike the dilute suspension-limited Rotne-Prager-Yamakawa (RPY) tensor that neglects the higher-order many-body effects \cite{RPY_JFM2014}, accounts for many-body HIs by combining far-field multipole expansions with near-field lubrication models. SD initially constructs the mobility matrix by pairwise summing far-field HIs using multipole expansions, truncated at the stresslet level. The many-body nature of far-field HI is then addressed by inverting this mobility matrix, resulting in the far-field resistance matrix \cite{Durlofsky1987SD}. Subsequently, near-field lubrication effects are added through pairwise asymptotic resistance functions \cite{Kim1985exact_resistance3}. The final mobility matrix, obtained by inverting the combined resistance matrix, allows for the calculation of particle velocities given external forces exerted on each particle. The original SD implementation's $O(N^2)$ matrix construction and $O(N^3)$ matrix inversion complexity for $N$ particles led to the development of Accelerated Stokesian Dynamics (ASD) \cite{sierou2001ASD,WANG2016ASD2,Ouaknin2021ParaASD}. ASD reduces complexity to $O(N \log N)$ by employing Ewald sum splitting and fast Fourier transforms, efficiently handling far-field HIs.

Distinct from these traditional approaches, machine learning methods extract relevant information from data to construct computationally efficient surrogate models, thereby accelerating simulations. The hierarchical machine learning approach presented by Siddani and Balachandar \cite{Siddani2023trinary} enables efficient prediction of neighbor-induced force and torque fluctuations in viscous fluid at Reynolds numbers below a few hundred, with improved accuracy achieved by including three-body interactions. This method, however, was only demonstrated for modeling small systems of particles. Graph neural networks (GNNs) show promise for inferring the dynamics of arbitrary numbers of particles \cite{Battaglia2016Interaction,Li2019Propagation,Sanchez-Gonzalez2020GNS,Bapst2020GNN}. However, these previously reported GNN-based methodologies are unsuitable for modeling particulate suspensions due to high training costs and the inability to capture long-range interactions, as modeling particulate suspensions requires accounting for significant long-range HIs and is limited by the impracticality of generating vast training data. 

In our prior work, we introduced the Hydrodynamic Interaction Graph Neural Network (HIGNN) \cite{Ma2022HIGNN}, a new framework that harnesses the power of GNNs to enable fast simulation of particulate suspensions. This framework has subsequently been utilized to examine a variety of suspension drop behaviors, including shape deformation, disintegration, and coalescence \cite{ma2024shape}. The basic idea of HIGNN is to represent a suspension of particles as a graph, where the particles and two-body interactions are represented as \textit{vertices} and \textit{edges}, respectively. Besides, to incorporate HIs beyond two-body, i.e., $m$-body interactions with $m > 2$, the graph is generalized to include higher-order connectivity. For example, a \textit{3-clique} connectivity between three vertices (particles) is introduced into the graph and used to describe three-body connections and thereby to capture the three-body HI effects. By introducing higher-order connectivity in the graph and utilizing corresponding convolutional operations, HIGNN enables accurately modeling many-body HIs between particles. As discussed, each connectivity within the graph, such as edge or 3-clique, accounts for a corresponding type of $m$-body interactions (two-body or three-body, respectively). The GNN learns the mapping from these connectivities to their associated interactions. This direct correspondence between graph connectivity and physical interactions lends \textit{explainability} to the HIGNN model, distinguishing it from ``black-box" neural network models. Another notable advantage of HIGNN is its \textit{transferability} across different systems. Similar to SD, it directly models the mobility tensor, and hence can be applied to simulating suspensions of particles subject to any type and magnitudes of external forces. The dependence of each $m$-body (e.g., two-body or three-body) interactions on the relative positions of associated particles remains invariant, irrespective of the total number or concentration of particles in the suspension. Therefore, once the HIGNN is trained for a domain, it can be applied to simulating suspensions of varying numbers of particles in the same domain. Furthermore, training HIGNN requires minimal computational resources: if achieving the desired accuracy necessitates preserving $m$-body HIs, the training process only requires data for up to $m$ particles, sampled across a range of spatial configurations within the domain of interest. HIGNN offers the potential for significant computational speedup by requiring only a forward pass through its neural networks at each time step, unlike traditional methodologies discussed above \cite{IBM_CMAME2016,ParticulateBI_JCP2020, YEO2010FCM3, GMLS_ScalableSolver_Ye2022, Brady1988_SDreview,  Ouaknin2021ParaASD} that typically require solving large-scale linear systems. However, HIGNN's overall prediction cost scales quadratically, $O(N^2)$, exceeding the $O(N\log N)$ scaling of ASD \cite{Ouaknin2021ParaASD}. As a result, while HIGNN outperforms ASD for small- to moderate-scale problems ($N \le 10^4$) \cite{Ma2022HIGNN,ma2024shape}, its cost approaches or surpasses ASD for very large-scale simulations. This $O(N^2)$ complexity stems from the slow decay of two-body HIs, necessitating edges between all particle pairs and a fully-connected graph representation of the particle system. This results in an $O(N^2)$ time complexity for the matrix-vector multiplication during inference. 

To ensure consistent efficiency superiority across all scales, in the present work, we aim to reduce the prediction cost scaling of HIGNN to quasi-linear, i.e., $O(N\log N)$. To achieve this, we propose integrating the hierarchical matrix ($\mathcal{H}$-matrix) techniques \cite{Hackbusch2015Hmatrix, Bebendorf2008Hmatrix} into the HIGNN framework. The core principle of $\mathcal{H}$-matrix involves partitioning a matrix into blocks. The partitioned blocks meeting the admissibility condition \cite{Bebendorf2008Hmatrix} can then be approximated by low-rank matrices \cite{Tyrtyshnikov1996cross_proof1,Tyrtyshnikov2000cross_proof2}. This low-rank approximation can significantly reduce the cost of matrix-vector multiplication. Constructing an $\mathcal{H}$-matrix requires $O(N\log N)$ operations, while matrix-vector multiplication with an $\mathcal{H}$-matrix typically exhibits $O(N)$ complexity. Therefore, integrating $\mathcal{H}$-matrix addresses our need to reduce HIGNN's computational complexity to quasi-linear. Moreover, as shown later in our numerical tests, the actual computing time associated with $\mathcal{H}$-matrix remains low. Consequently, even for simulations involving very large-scale particulate suspensions (e.g., containing more than millions of particles), computational costs related to hardware resources and wall time can be maintained at a low level. It is worth noting that our approach differs from previous work incorporating hierarchical structures into GNNs, such as Multipole Graph Neural Operator \cite{Li2020MultipoleGNN} and Hierarchical Message-Passing Graph Neural Networks \cite{Zhong2023HierarchicalGNN}. These prior methods formulate the entire model structure as a $\mathcal{H}$-matrix and learn the message passing between different graph levels (i.e. low-rank approximation in $\mathcal{H}$-matrices) from data. Thus, these approaches require training datasets containing large numbers of particles, which is not feasible in our context due to the high cost of data generation. Instead, we preserve HIGNN's advantage of training on data with small numbers of particles, while only applying $\mathcal{H}$-matrix techniques during the inference stage to build our $\mathcal{H}$-HIGNN framework.

Integrating $\mathcal{H}$-matrices into the HIGNN framework requires a concerted technical effort, involving several key components. First, recognizing the inherent geometric structure of the particle system, we performed cluster-tree block partitioning \cite{Hackbusch1989cluster} by using the geometric information embedded in the graph. Second, the admissibility is determined based on the relative distance between clusters and their respective sizes. Third, by leveraging this admissibility condition and the asymptotic smoothness of two-body HIs, we implemented an adaptive cross approximation (ACA) algorithm \cite{Tyrtyshnikov2000cross_proof2} to achieve effective low-rank approximations for a significant portion of partitioned blocks in the mobility matrix. Fourth, unlike conventional $\mathcal{H}$-matrices, where entries are typically scalars, each element in our mobility matrix is a $3 \times 3$ submatrix. This necessitated the development of specialized matrix operations and data structures in both the ACA algorithm and code implementation to efficiently handle these submatrices within the $\mathcal{H}$-matrix framework, ensuring compatibility and performance. Finally, through careful optimization of GPU latency, workload distribution, and inter-GPU communication, coupled with the Kokkos programming model \cite{9485033}, our implementation achieves a computer wall-time that closely approximates the theoretical $O(N\log N)$ scaling and yields a near-ideal strong scalability for parallel efficiency. Such $\mathcal{H}$-matrix accelerated HIGNN is designated as $\mathcal{H}$-HIGNN. Through a comprehensive evaluation, we assess the accuracy of $\mathcal{H}$-HIGNN against the original HIGNN, and validate its computational complexity and scalability. Furthermore, we demonstrate its efficiency in simulating practically large-scale particle suspensions. By incorporating elastic bonding and bending forces between particles, $\mathcal{H}$-HIGNN can also be applied to simulating large-scale suspensions of flexible filaments.

The remainder of this paper is organized as follows. Section \ref{sec:HIGNN} provides a concise overview of the HIGNN framework. Section \ref{sec:H-matrix} details the $\mathcal{H}$-matrix methodology and the approach to integrate $\mathcal{H}$-matrix into HIGNN, including specific algorithms and the strategies for scalable implementation. Section \ref{sec:results} presents all results, including the assessment of $\mathcal{H}$-HIGNN's accuracy, computational complexity, and scalability, as well as its efficacy in simulating large-scale suspensions of particles and filaments. Finally, Section \ref{sec:conclusion} summarizes our work and highlights the key contributions.

\section{Overview of HIGNN}\label{sec:HIGNN}
To lay the groundwork for this study, this section provides a brief overview of the HIGNN framework introduced in our prior work \cite{Ma2022HIGNN,ma2024shape}. We begin by outlining the problem setup before reviewing the key aspects of HIGNN, including its theoretical foundation and framework structure. A discussion on the computational complexity and scaling properties of HIGNN's prediction cost then motivates the present study.

\subsection{Problem setup}
\label{subsec:problem_setup}
Consider a system of $N$ identical rigid spherical particles of radius $a$ suspended in an incompressible Newtonian fluid. Under the assumption of low Reynolds number ($Re =\rho Ua/\mu \ll 1$, where $\rho$ represents density, $\mu$ represents viscosity, and $U$ represents the characteristic particle velocity), the motion of the particles is governed by over-damped dynamics, neglecting inertia, as given by:
\begin{equation}\label{equ:BD_mobility}
\begin{bmatrix}
    \dot{\textbf{X}} \\
   \dot{\boldsymbol{\theta}}
\end{bmatrix}=
\begin{bmatrix}
    \mathbf{U} \\
    \boldsymbol{\omega}
\end{bmatrix}
= \textbf{M} (\textbf{X})\cdot 
\begin{bmatrix}
    \textbf{F} \\
    \textbf{T}
\end{bmatrix}\;,
\end{equation}
where $\textbf{X} \in \mathbb{R}^{3N}$ and $\boldsymbol{\theta} \in \mathbb{R}^{3N}$ denote the particles' spatial positions and orientations, respectively; $\mathbf{U}\in \mathbb{R}^{3N}$ and $\boldsymbol{\omega}\in \mathbb{R}^{3N}$ denote the translational and rotational velocities of particles, respectively; $\textbf{F} \in \mathbb{R}^{3N}$ and $\textbf{T} \in \mathbb{R}^{3N}$ encompass all external forces and torques applied on each particle and may be attributed to a variety of sources, including gravity, external fields, and interparticle potentials. Here, $\textbf{M}(\textbf{X})\in\mathbb{R}^{6N\times 6N}$ is the mobility tensor (matrix), where each entity $\mathbf{M}_{ij} \in \mathbb{R}^{3 \times 3} $ quantifies the translational or rotational velocity imparted to particle $i$ per unit external force or torque applied on particle $j$. Due to HI, each particle's mobility is affected by the presence of others, hence $\textbf{M}$ must account for many-body effects and depends on the configuration $\textbf{X}$ of all particles in the suspension. The key to simulating particulate suspensions, i.e., evolving the dynamical equation \eqref{equ:BD_mobility}, is the calculation of the mobility matrix $\textbf{M}$ at each time step.

Two types of traditional approaches exist for obtaining the mobility matrix $\textbf{M}(\textbf{X})$: directly solving the Stokes equations with prescribed boundary conditions on the particles' surfaces  \cite{ParticulateReview_Maxey2017,YEO2010FCM3,GMLS_StokesColloid_HuCMAME2019,GMLS_ScalableSolver_Ye2022}; and approximating $\textbf{M}(\textbf{X})$ through multipole expansions truncated at the stresslet level, as in SD and its variants \cite{Brady1988_SDreview,sierou2001ASD,Ouaknin2021ParaASD,Fiore2019FSD}. However, both methods are computationally demanding, and the computed $\textbf{M}(\textbf{X})$ lacks transferability to other suspensions even with the same type of particles but different particle numbers or concentrations. To mitigate these limitations, we have proposed the HIGNN framework \cite{Ma2022HIGNN} that enables fast simulations and transferability across systems with different numbers of particles in the same domain. 

\subsection{Theoretical foundation}\label{subsec:theory}
Based on the many-body nature of HI \cite{Mazur1982ManysphereHI} and our analysis \cite{Ma2022HIGNN}, the mobility tensor $\textbf{M}(\textbf{X})$ can be expressed as an expansion of additive terms arising from single-body and different $m$-body contributions, e.g., two-body, three-body, etc. From theoretical analysis on the decaying orders of different $m$-body contributions with respect to the characteristic particles' separation distance (denoted as $r$) \cite{Mazur1982ManysphereHI,Ma2022HIGNN}, we know that for greater $m$, the corresponding $m$-body contribution decays at a faster rate in terms of $r$. In the original HIGNN framework \cite{Ma2022HIGNN}, the $m$-body contributions with $m \ge 4$ are truncated, retaining only two- and three-body interactions in HI. While used for proof of concept, this approximation is not inherent to the framework and has been validated for suspensions that are not too concentrated, achieving the desired level of accuracy \cite{Ma2022HIGNN,ma2024shape}. Henceforth, the velocity of each particle can be approximated as:
\begin{equation}\label{Eq:theo_basis_trun_final}
\begin{split}
    \mathbf{U}_i  & \approx  \boldsymbol{\alpha}_1 (\mathbf{X}_i) \cdot \mathbf{F}_i + \sum\limits_{\substack{{j = 1} \\ j\neq i}}^N \boldsymbol{\alpha}_2 (\mathbf{X}_i, \mathbf{X}_j)  \cdot \mathbf{F}_{i,j}  + \sum\limits_{\substack{ j, k: j \neq i,k \neq i \\i,j \in \mathcal{N}(k) }}^N \boldsymbol{\alpha}_3 (\mathbf{X}_i, \mathbf{X}_k, \mathbf{X}_j ) \cdot \mathbf{F}_{i,j} \;,
\end{split}
\end{equation}
where $\mathbf{F}_{i,j} = [\mathbf{F}_{i}^T , \mathbf{F}_{j}^T ]^T \in \mathbb{R}^{6 \times 1}$; and $\mathcal{N}(*)$ denotes the neighbors of particle $*$ within certain distance. $\boldsymbol{\alpha}_1(\mathbf{X}_i)$ denotes the general single-particle mobility in the absence of HI between particles, with $\boldsymbol{\alpha}_1(\mathbf{X}_i)=(6\pi \mu a)^{-1} \mathbf{I}$ for spherical particles in an unbounded domain. $\boldsymbol{\alpha}_2 = [ \boldsymbol{\alpha}_2^{(s)},  \boldsymbol{\alpha}_2^{(t)}] \in \mathbb{R}^{3 \times 6}$ and $\boldsymbol{\alpha}_3 = [ \boldsymbol{\alpha}_3^{(s)}, \boldsymbol{\alpha}_3^{(t)}] \in \mathbb{R}^{3 \times 6} $ denote the two-body and three-body HI, respectively. Here, $\boldsymbol{\alpha}^{(s)}_m$ and $\boldsymbol{\alpha}^{(t)}_m$ can be regarded as the $m$-body HI contributions to the diagonal ($\mathbf{M}_{ii}$) and off-diagonal ($\mathbf{M}_{ij}$, $i\neq j$) elements of the mobility matrix, respectively. For example, $\boldsymbol{\alpha}_3^{(s)} $ corresponds to the contribution of three-body HI to $\mathbf{M}_{ii}$, quantifying the effect on the mobility of particle $i$ due to the simultaneous presence of particles $j$ and $k$. From our analysis \cite{Ma2022HIGNN}, for spherical particles in unbounded domain, the leading orders of $\boldsymbol{\alpha}^{(t)}_2$, $\boldsymbol{\alpha}^{(s)}_2$, $\boldsymbol{\alpha}^{(t)}_3$, and $\boldsymbol{\alpha}^{(s)}_3$ with respect to $r$ are $O(r^{-1})$, $O(r^{-4}) $, $O(r^{-4}) $, and $O(r^{-7}) $, respectively. To reduce computational cost, we exploit the fact that $\boldsymbol{\alpha}^{(s)}_2$, $\boldsymbol{\alpha}^{(t)}_3$, and $\boldsymbol{\alpha}^{(s)}_3$ decay much faster than $\boldsymbol{\alpha}^{(t)}_2$ and hence set a cutoff distance $R_\text{cut}$ in the way that only the neighbors of a particle within $R_\text{cut}$ are included in the calculations of $\boldsymbol{\alpha}^{(s)}_2$,  $\boldsymbol{\alpha}^{(t)}_3$, and $\boldsymbol{\alpha}^{(s)}_3$, while all neighbors are considered for the calculation of  $\boldsymbol{\alpha}^{(t)}_2$.   

Although the present work focuses on particulate suspensions in unbounded domains to facilitate the discussion on prediction cost and scalability—as unbounded domains readily accommodate arbitrarily large numbers of particles—it is worth noting that Eq. \eqref{Eq:theo_basis_trun_final} is applicable to general scenarios, including unbounded, periodic, and bounded domains. For different types of domains, the functions for $\boldsymbol{\alpha}_1$, $\boldsymbol{\alpha}_2$, and $\boldsymbol{\alpha}_3$ would be different. Moreover, while we limit our discussion herein to translational motions of spherical particles without torques, HIGNN can be straightforwardly extended to include rotational velocities and torques (see \cite{Ma2022HIGNN} for a detailed discussion and derivation).

\subsection{Framework structure}
Based on Eq.~\eqref{Eq:theo_basis_trun_final}, HIGNN is constructed by introducing graph data structure to characterize $m$-body connections and modeling the nonlinear functions $\boldsymbol{\alpha}_m$ as learnable neural networks. For a suspension of $N$ particles, we build a graph of $N$ vertices, denoted as $\boldsymbol{\mathcal{V}} = [1,2,\dots,N]^T$, with each particle regarded as a vertex. Each vertex is associated with a \textit{feature vector} $\mathbf{X}_i$ and an output vector $\mathbf{U}_i$. Between any two vertices, a directed edge $(j, i)$ is defined, pointing from the source vertex $j$ to the target vertex $i$, to characterize a two-body connection. The total number of edges is $N_e = N(N-1)$, to encompass all two-body connections and their HIs. All the edges constructed are stored in matrix $\boldsymbol{\mathcal{E}} \in \mathbb{Z}^{2 \times N_e}$, where the two entities of each column correspond to the indices of the source and target vertices of an edge, respectively. The \textit{edge feature} is defined by the external forces exerted on the target and source vertices (particles), i.e., $\mathbf{F}_{i,j}$ in Eq.~\eqref{Eq:theo_basis_trun_final}. HIGNN further introduces ``{3-clique}" connectivity in the graph to characterize three-body HIs. Each {3-clique}, denoted by $(j, k, i)$, is also directed and characterizes how the presence of vertex (particle) k modifies the impact of $\mathbf{F}_i$ and $\mathbf{F}_j$ on the velocity of vertex (particle) $i$. A {3-clique} is built only when the three vertices are within the preset $R_\text{cut}$, i.e., {3-clique}$(j,k,i)$ exists only if $i, j \in \mathcal{N}(k)$. Such defined {3-cliques} are represented by matrix $\boldsymbol{\mathcal{F}} \in \mathbb{Z}^{3 \times N_f}$ with $N_f$ the total number of {3-cliques}, and the entities in each column of $\boldsymbol{\mathcal{F}}$ correspond to the indices of the three vertices forming a {3-clique}. Same as the edge feature, the \textit{{3-clique} feature} is defined by the external forces exerted on the target and source vertices, i.e., $\mathbf{F}_{i,j}$ in Eq.~\eqref{Eq:theo_basis_trun_final}. Putting together, a graph $ \boldsymbol{\mathcal{G} = (\mathcal{V}, \mathcal{E}, \mathcal{F})}$, consisting of vertices, edges, and {3-cliques}, fully characterizes the entire suspension system. 

\begin{figure}
\centering
\begin{subfigure}{1.0\textwidth}
\centering
\includegraphics[width=0.8\textwidth]{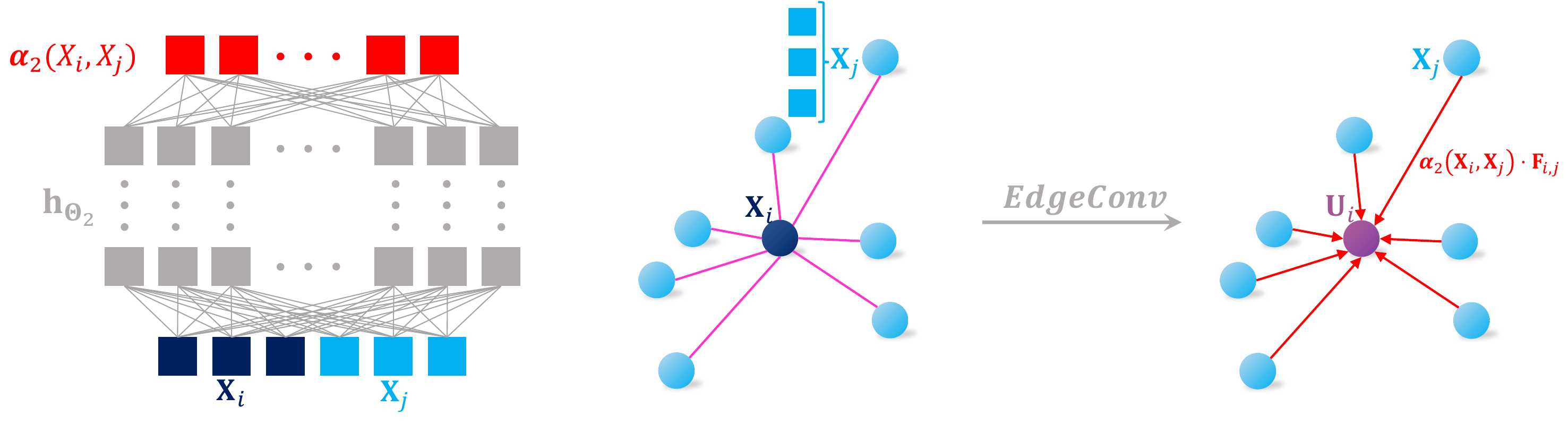}
\caption{The surrogate model of $\boldsymbol{\alpha}_2(\mathbf{X}_j, \mathbf{X}_i)$ is the MLP neural network $\mathbf{h}_{\boldsymbol{\Theta}_2}$ obtained through EdgeConv.} 
\label{sfig:edge_conv}
\end{subfigure}

\vspace{0.8cm}
\begin{subfigure}{1.0\textwidth}
\centering
\includegraphics[width=0.8\textwidth]{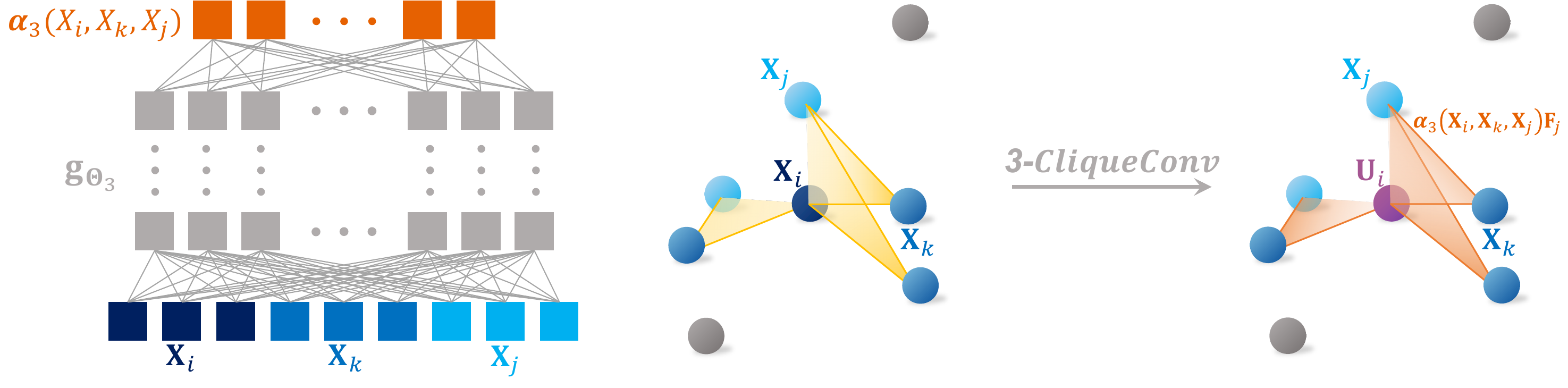}
\caption{The surrogate model of $\boldsymbol{\alpha}_3(\mathbf{X}_j, \mathbf{X}_k, \mathbf{X}_i)$ is the MLP neural network  $\mathbf{g}_{\boldsymbol{\Theta}_3}$ obtained through 3-CliqueConv.} 
\label{sfig:face_conv}
\end{subfigure}
\caption{Main components of the HIGNN framework.} 
\label{fig:graph_conv}
\end{figure}

In addition, HIGNN constructs two Multilayer Perceptron (MLP) neural networks $\mathbf{h}_{\boldsymbol{\Theta}_2}$ and $ \mathbf{g}_{\boldsymbol{\Theta}_3}$ on edge and 3-clique connections as the surrogates of $\boldsymbol{\alpha}_2$ and $\boldsymbol{\alpha}_3$, respectively, with $\boldsymbol{\Theta}_2$ and $\boldsymbol{\Theta}_3$ the network parameters. To this end, the information on all edges and {3-cliques} associated with the target vertex are aggregated by edge convolutional (EdgeConv) \cite{wang2019edge_conv} and {3-clique} convolutional ({3-Clique}Conv, referred to as FaceConv in \cite{Ma2022HIGNN,ma2024shape}) operations, respectively. The main components of the HIGNN framework are summarized in FIG. \ref{fig:graph_conv}. 
Finally, the resulting surrogate model by HIGNN for predicting the velocity of a particle can be expressed as:
\begin{equation}\label{Eq:U_surrogate_3body}
\begin{split}
    \mathbf{U}_i^{\text{HIGNN}}  & = \boldsymbol{\alpha}_1 (\mathbf{X}_i) \cdot \mathbf{F}_i + \sum\limits_{\substack{j \\(j,i)\in \boldsymbol{\mathcal{E}}}}
    \mathbf{h}_{\boldsymbol{\Theta}_2} (\mathbf{X}_i, \mathbf{X}_j) \cdot \mathbf{F}_{i,j}  +
    \sum \limits_{\substack{j,k \\(j,k,i) \in \boldsymbol{\mathcal{F}}}}
    \mathbf{g}_{\boldsymbol{\Theta}_3} (\mathbf{X}_i, \mathbf{X}_k, \mathbf{X}_j) \cdot \mathbf{F}_{i,j} 
\end{split}
\end{equation}

\subsection{Training}
\label{subsec:Training_inference_HIGNN}
The training of HIGNN requires minimal computational resources, as discussed below. Training HIGNN to determine $\mathbf{h}_{\boldsymbol{\Theta}_2}$ and $ \mathbf{g}_{\boldsymbol{\Theta}_3}$, for use in Eq. \eqref{Eq:U_surrogate_3body}, requires only data of two or three particles sampled with different spatial configurations in the domain of interest. In general, if the attainment of desired accuracy necessitates the retention of $m$-body contributions in the mobility matrix $\textbf{M}(\textbf{X})$, the training process requires data comprising up to $m$ particles, sampled across a range of spatial configurations within the domain. For bounded domains, we need to train another neural network as the surrogate for $\boldsymbol{\alpha}_1 (\mathbf{X}_i)$, using the data of a single particle at different spatial positions in the domain. However, for unbounded and periodic domains, $\boldsymbol{\alpha}_1$ admits an analytical solution. Also, 
for particles in unbounded or periodic domains, HIs only depend on their relative positions or interparticle distances. Thus, normalizing the neural network outputs by the corresponding interparticle distances can improve the training performance. Eq. \eqref{Eq:U_surrogate_3body} can then be rewritten as:
\begin{equation}\label{Eq:3body_GNN_relative}
\begin{split}
    \mathbf{U}^{\text{HIGNN}}_i & = \boldsymbol{\alpha}_1 \cdot \mathbf{F}_i + \sum\limits_{\substack{j :(j,i)\in \boldsymbol{\mathcal{E}}}}  \frac{ \hat{\mathbf{h}}_{\boldsymbol{\Theta}_2}  ( \mathbf{X}_j - \mathbf{X}_i)}{\| \mathbf{X}_j - \mathbf{X}_i \|_2 } \cdot \mathbf{F}_{i,j} + \sum\limits_{j,k: (j,k,i) \in \boldsymbol{\mathcal{F}}}  \frac{\hat{\mathbf{g}}_{\boldsymbol{\Theta}_3} (\mathbf{X}_j - \mathbf{X}_i, \mathbf{X}_k- \mathbf{X}_i) }{ \| \mathbf{X}_j - \mathbf{X}_k \|_2 \| \mathbf{X}_k - \mathbf{X}_i \|_2 } \cdot \mathbf{F}_{i,j} \;,
\end{split} 
\end{equation}
where $ \| \cdot \|_2 $ denotes the 2-norm of a vector. $\hat{\mathbf{h}}_{\boldsymbol{\Theta}_2} $ and $\hat{\mathbf{g}}_{\boldsymbol{\Theta}_3}$ are correlated with $\mathbf{h}_{\boldsymbol{\Theta}_2} $ and $\mathbf{g}_{\boldsymbol{\Theta}_3}$ in Eq.~\eqref{Eq:U_surrogate_3body} by $ \mathbf{h}_{\boldsymbol{\Theta}_2} = \frac{ \hat{\mathbf{h}}_{\boldsymbol{\Theta}_2}  ( \mathbf{X}_j - \mathbf{X}_i)}{\| \mathbf{X}_j - \mathbf{X}_i \|_2 }$ and $ \mathbf{g}_{\boldsymbol{\Theta}_3} = \frac{\hat{\mathbf{g}}_{\boldsymbol{\Theta}_3} (\mathbf{X}_j - \mathbf{X}_i, \mathbf{X}_k- \mathbf{X}_i) }{ \| \mathbf{X}_j - \mathbf{X}_k \|_2 \| \mathbf{X}_k - \mathbf{X}_i \|_2 } $, respectively. 

In the generation of data involving two particles, the maximum distance considered between the two particles was set sufficiently large (e.g., 1000) so as to encompass the range of distances likely to be encountered in subsequent tests and simulations. All numerical values are non-dimensional, with the radius of particles normalized to 1. For the data involving three particles, used exclusively to train the three-body HIs, the maximum interparticle distance was set to the cutoff distance, $ R_\text{cut}$. Our prior studies \cite{Ma2022HIGNN} have determined $R_\text{cut} = 5.0$, and hence this cutoff distance is used to define {3-clique} connectivity in the graph. For each dataset, particle positions were randomly generated within the established distance constraints, and each particle was assigned a unit force randomly sampled from the set: $\mathbf{F} = [1, 0, 0]^T$, $[0,1,0]^T$, or $[0,0, 1]^T$. For the purposes of the present study, 100,000 distinct configurations of two particles and 50,000 distinct configurations of three particles were generated in an unbounded domain. For each configuration, a numerical solver can be invoked to calculate particle velocities for a single time step, given the particle positions and assigned forces. Due to the code accessibility, we used SD \cite{SD_opensource}. Each set of particle positions, applied forces, and resulting velocities constitutes a single data point to be utilized in the subsequent training and validation processes. The generation of each data point only requires a wall time of 0.014 seconds on a workstation equipped with an Intel(R) Xeon(R) E5-2698 v4 CPU @ 2.20GHz. 

Both $\hat{\mathbf{h}}_{\boldsymbol{\Theta}_2} $ and $\hat{\mathbf{g}}_{\boldsymbol{\Theta}_3}$ are MLP neural networks, each with three hidden layers of 128, 512, and 128 nodes, respectively, and a hyperbolic tangent (tanh) activation function. The loss function used for training is defined as:
\begin{equation}\label{Eq:Loss}
    \mathcal{L} = \frac{1}{N_\text{train}} \left. \sum\limits_{i=1}^{N_\text{train}} \left( \|  \mathbf{U}_i - \mathbf{U}^\text{HIGNN}_i \|_2^2  \middle /  \| \mathbf{U}_i \|_2^2\right) \right.\;.
\end{equation}
Using $N_\text{train}=120,000$ data for training (with the remainder reserved for testing and validation), the training process requires only $\sim$32 min on a desktop with an NVIDIA RTX 3070 GPU and an AMD 3900XT CPU @ 3.79 GHz.

\subsection{Prediction: transferability and cost}
As HIGNN essentially models the mobility tensor, the trained HIGNN is transferable to predicting the velocities of particles (suspended in fluid) subject to any types and magnitudes of external forces. Moreover, all $\boldsymbol{\alpha}_i$ terms in $\textbf{M}(\textbf{X})$ are invariant regardless of the number of particles, and hence the trained surrogate models for $\boldsymbol{\alpha}_i$ in HIGNN are transferable across suspensions of different concentrations or numbers of particles in the same domain. Because of its inherent transferability, the trained HIGNN eliminates the need for retraining when simulating particulate suspensions of varying concentrations and particle numbers under arbitrary external forces.

To predict the time evolution of particles, each particle's position is updated from its velocity by numerically integrating $ \dot{\textbf{X}} =\mathbf{U}$ over time. At each time step, the trained HIGNN is called to predict all particles' velocities, using the graph $ \boldsymbol{\mathcal{G} = (\mathcal{V}, \mathcal{E}, \mathcal{F})}$ , which fully characterizes the particle system, as input. This allows us to effectively simulate the dynamics of particles in suspension.

HIGNN can potentially achieve significant computational speedup by requiring only a forward pass through its MLP neural networks at each time step, in contrast to traditional simulation methodologies \cite{IBM_CMAME2016,ParticulateBI_JCP2020, YEO2010FCM3, GMLS_ScalableSolver_Ye2022, Brady1988_SDreview,  Ouaknin2021ParaASD}, which usually require solving a large-scale linear system at each time step. The computational complexity of HIGNN is directly related to the number of edges and {3-cliques} in the graph, as neural network inferences are performed on each edge and {3-clique}. As discussed above, the two-body HI not only dominates the short-range lubrication effect but also decays very slowly ($\sim O(r^{-1})$) in long range. Thus, for each particle, we need to consider all other particles as its neighbors  for the calculation of  $\boldsymbol{\alpha}^{(t)}_2$. In HIGNN, this translates to the construction of edges between any two particles, leading to a fully-connected graph characterized by $N(N-1)$ edges. As for the {3-cliques}, which characterize three-body connections, the faster decaying rate of three-body HIs allows for the implementation of a finite cutoff distance ($R_\text{cut}$), thereby limiting the number of {3-cliques} to $O(N)$. Therefore, the overall computational complexity (prediction cost) of HIGNN scales as $O(N^2)$ with $N$ the total number of particles, which is greater than the scaling $O(N\log N)$ of ASD \cite{Ouaknin2021ParaASD}. This implies that while HIGNN's computational cost is significantly lower than ASD for small- to moderate-scale problems (as shown in our previous work \cite{Ma2022HIGNN,ma2024shape}), for simulations of very large-scale particulate suspensions, HIGNN's cost can approach or surpass that of ASD. Therefore, to achieve a consistent superiority on computational efficiency, in the present work we aim to reduce the scaling of the prediction cost of HIGNN to quasi-linear, i.e., ${O}(N\log N)$. By such, even for simulating very large-scale particulate suspensions (e.g., containing more than millions of particles), the computational cost can still be maintained low. For achieving that, we propose to integrate the $\mathcal{H}$-matrix techniques \cite{Hackbusch2015Hmatrix, Bebendorf2008Hmatrix} into the HIGNN framework. The detailed methodology is introduced in the next section.

\section{Hierarchical Matrix ($\mathcal{H}$-Matrix)$-$Accelerated HIGNN}
\label{sec:H-matrix}

To facilitate the discussion, Eq.~\eqref{Eq:theo_basis_trun_final} is decomposed and rewritten as a summation of matrix-vector multiplications:
\begin{equation}\label{Eq:theo_basis_mat}
    \mathbf{U}  \approx  \mathbf{M}^{(1)} \cdot \mathbf{F} + \mathbf{M}^{(t, 2)} \cdot \mathbf{F} + \mathbf{M}^{(s, 2)} \cdot \mathbf{F} + \mathbf{M}^{(t, 3)} \cdot \mathbf{F} + \mathbf{M}^{(s, 3)} \cdot \mathbf{F} \;,
\end{equation}
where $\mathbf{M}^{(1)} $, $\mathbf{M}^{(s, 2)}$ and $\mathbf{M}^{(s, 3)}$ are all diagonal matrices with each element (a $3\times 3$ submatrix) calculated from $\boldsymbol{\alpha}_1$, $\boldsymbol{\alpha}^{(s)}_2$ and $\boldsymbol{\alpha}^{(s)}_3$, respectively; 
the matrix $\mathbf{M}^{(t, 2)} $ is constructed from the two-body interactions $\boldsymbol{\alpha}_2^{(t)}$, and  $\mathbf{M}^{(t, 3)} $ from the three-body interactions $\boldsymbol{\alpha}_3^{(t)}$. Following the preceding analysis, the matrix-vector multiplications appearing in Eq. \eqref{Eq:theo_basis_mat} typically exhibit $O(N)$ complexity due to the sparsity of the associated matrices, but with the exception of the multiplication $\mathbf{M}^{(t, 2)} \cdot \mathbf{F}$, which has $O(N^2)$ complexity because $\mathbf{M}^{(t, 2)}$ is a dense matrix. Thus, we integrate the $\mathcal{H}$-matrix techniques \cite{Hackbusch2015Hmatrix, Bebendorf2008Hmatrix} to reduce the computational complexity for the multiplication $\mathbf{M}^{(t, 2)} \cdot \mathbf{F}$ from $O(N^2)$ to $O(N\log N)$. Note that $\mathbf{M}^{(t, 2)}$ is in fact assembled by $N \times N $ submatrices, where each submatrix {$\mathbf{M}^{(t, 2)}_{ij} \in \mathbb{R}^{3\times 3} $ corresponds to $\boldsymbol \alpha_2^{(t)} (\mathbf{X}_i, \mathbf{X}_j)$. Following the convention of $\mathcal{H}$-matrix, it is more convenient to let the matrix's dimension equal to the number of particles. Hence, we denote each submatrix $\mathbf{M}^{(t, 2)}_{ij}$ as an element or entry of the matrix, and thereby $\mathbf{M}^{(t, 2)}$ is referred to as an $N \times N$ matrix.\footnote{For the subsequent discussion, it should be noted that when performing any operation on elements, each element corresponds to a $3 \times 3$ submatrix, rather than a scalar value.}

The basic idea of $\mathcal{H}$-matrix is to partition $\mathbf{M}^{(t, 2)}$ into blocks based on geometric information. The blocks satisfying the admissibility condition \cite{Bebendorf2008Hmatrix}, referred to as the admissible blocks, can then be approximated by low-rank matrices \cite{Tyrtyshnikov1996cross_proof1,Tyrtyshnikov2000cross_proof2}. Intuitively, an admissible block can be regarded as a block where the associated vertices exhibit relatively longer edge connections. A more rigorous mathematical definition is given in \S\ref{subsec:block-partition}. After constructing the low-rank approximation, the complexity of the matrix-vector multiplication would be effectively reduced. For example, considering two subsets of vertices (or particles), one with $c$ particles, and the other with $q$ particles, the block representing the two-body interactions between these two sets of particles is denoted as $ \mathbf{G} \in \mathbb{R}^{c \times q} $. Its $l$-rank approximation can be expressed as: $\mathbf{G} \approx \mathbf{C} \mathbf{Q}$, where $\mathbf{C} \in \mathbb{R}^{c \times l} $ and $\mathbf{Q} \in \mathbb{R}^{l \times q} $. When multiplying with a vector $\mathbf{b} \in \mathbb{R}^q$, we can first calculate the product $ \mathbf{Q}\mathbf{b} $ and then multiply the resultant vector with $\mathbf{C}$. By such, the number of operations required can be reduced from ${O}(cq) $ to ${O}((c+ q)l)$. Since the function $\boldsymbol{\alpha}_2^{(t)}$ is asymptotically smooth, $\mathbf{G}$ can be well approximated by the $l$-rank approximation with $l \ll c$ and $l\ll q$, as long as $\mathbf{G}$ satisfies the admissibility condition \cite{Bebendorf2008Hmatrix, Tyrtyshnikov1996cross_proof1, Tyrtyshnikov2000cross_proof2}. Thus, the cost of the matrix-vector multiplication can be significantly reduced by the low-rank approximation. Given a configuration $\mathbf{X}$ of $N$ particles, we can employ the cluster-tree partitioning algorithm \cite{Hackbusch1989cluster} to partition the associated matrix $\mathbf{M}^{(t, 2)}$ into blocks such that each block satisfying the admissibility condition can be well approximated by a low-rank approximation. The resulting matrix, composed largely of low-rank blocks, is the so-called $\mathcal{H}$-matrix. 

In this work, we employ the adaptive cross approximation (ACA) \cite{Tyrtyshnikov2000cross_proof2} to find a low-rank approximation, thereby avoiding the computation of all entries of the original matrix. As a result, constructing $\mathcal{H}$-matrix requires ${O}(N\log N)$ operations, whereas the matrix-vector multiplication with an $\mathcal{H}$-matrix exhibits ${O}(N)$ complexity. More details about the cluster-tree block partitioning and ACA are provided as follows.

\subsection{Cluster-tree block partitioning}\label{subsec:block-partition}
To build an $\mathcal{H}$-matrix, the first step is to partition the matrix into blocks, which are organized in a tree structure defined as follows \cite{Hackbusch1989cluster}.

\begin{definition}
$\mathcal{T}_\mathbf{I}$ is a cluster tree corresponding to an index set $\mathbf{I}$

(1) if $\mathbf{I}$ is the root of $\mathcal{T}_\mathbf{I}$; 

(2) if $\boldsymbol \tau \in \mathcal{T}_{\mathbf{I}}$ is not a leaf, then $\boldsymbol \tau$ is the disjoint union of its two sons $\boldsymbol \tau_1 \in \mathcal{T}_{\mathbf{I}} $ and $\boldsymbol \tau_2 \in \mathcal{T}_{\mathbf{I}} $;

(3) if $\boldsymbol \tau \in \mathcal{T}_{\mathbf{I}}$ is a leaf, then the number of indices in set $\boldsymbol \tau$ is smaller than a preset threshold $C_\text{leaf}$, i.e., $|\boldsymbol \tau| < C_\text{leaf}$.
\end{definition}

In our case, the whole set of target and source vertices should be $\mathbf{I} = \{1, 2, \dots, N \}$, i.e., include all particles. This is because, as previously discussed, the two-body HIs in $\boldsymbol \alpha_2^{(t)} (\mathbf{X}_i, \mathbf{X}_j)$, which represent the impact of the force on particle $j$ (source) on the velocity of particle $i$ (target), decay slowly ($\sim O(r^{-1})$) at long range. Thus, all particle pairs must be considered, even those far apart. For cluster tree construction, this work utilizes geometry-based partitioning via bounding boxes, depicted in Figure \ref{fig:split}. While Figure \ref{fig:split} illustrates this in 2D for simplicity, all computations and simulations are performed in 3D. 
\begin{figure}[htbp]
    \centering
    \includegraphics[width = 10cm]{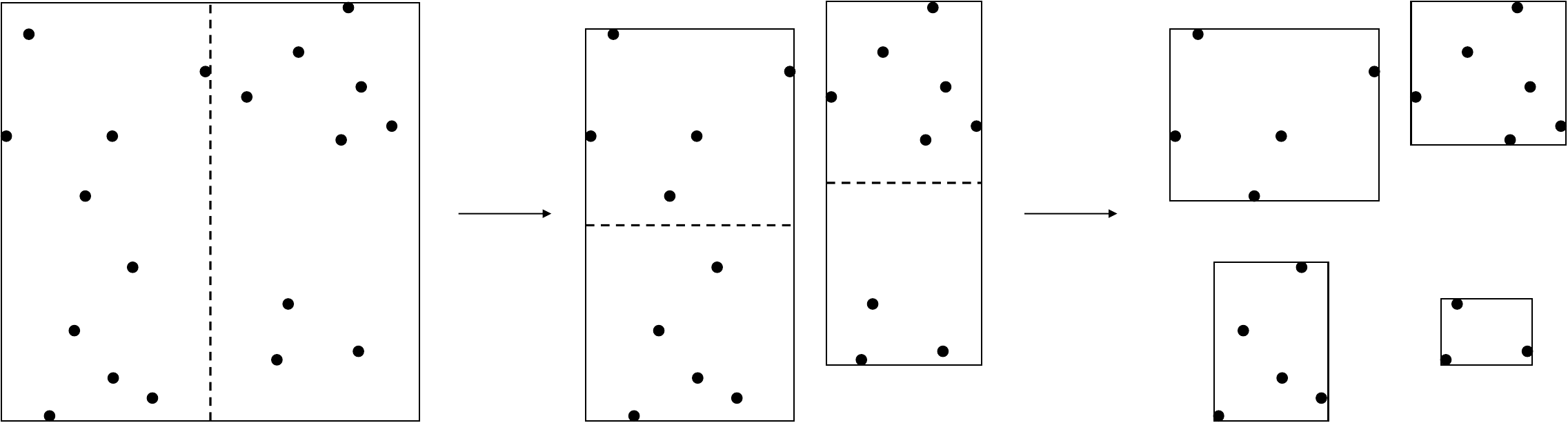}
    \caption{Geometry-based partitioning using bounding boxes, illustrated in 2D. The black dots represent particles. From left to right, the initial bounding box is first divided into two boxes along the $x$-direction, and then each of the new bounding boxes is divided along the $y$-direction.}
    \label{fig:split}
\end{figure}

Subsequent to block partitioning, we proceed to evaluate if a candidate block $(\boldsymbol \tau, \boldsymbol \sigma) \in \mathcal{T}_\mathbf{I} \times \mathcal{T}_\mathbf{I} $ satisfies the admissibility condition, thereby enabling a low-rank approximation.

\begin{definition}
A set of targets $\boldsymbol \tau$ and a set of sources $ \boldsymbol \sigma$ are called admissible \cite{Bebendorf2008Hmatrix}, if

\begin{equation}\label{Eq:admissibility}
    \max \{ \text{diam}(\boldsymbol \tau), \text{diam}( \boldsymbol \sigma) \} \leq \eta \text{dist} ( \boldsymbol \tau,  \boldsymbol \sigma) \;,\quad \text{with}~\eta > 0\;. 
\end{equation}
\end{definition}
Here, $\text{diam}(\cdot)$ denotes the diameter of a set (cluster), and $\text{dist} ( \boldsymbol \tau, \boldsymbol \sigma)$ represents the distance between $\boldsymbol \tau$ and $\boldsymbol \sigma$, as illustrated in Figure \ref{fig:diam_dist}(a). Due to the inherent complexity of determining the precise diameter of arbitrary clusters and the distance between them, we employ a practical approximation by calculating these values for the respective bounding boxes, as depicted in Figure \ref{fig:diam_dist}(b). 
\begin{figure}[htbp]
    \centering

    \includegraphics[width = 10cm]{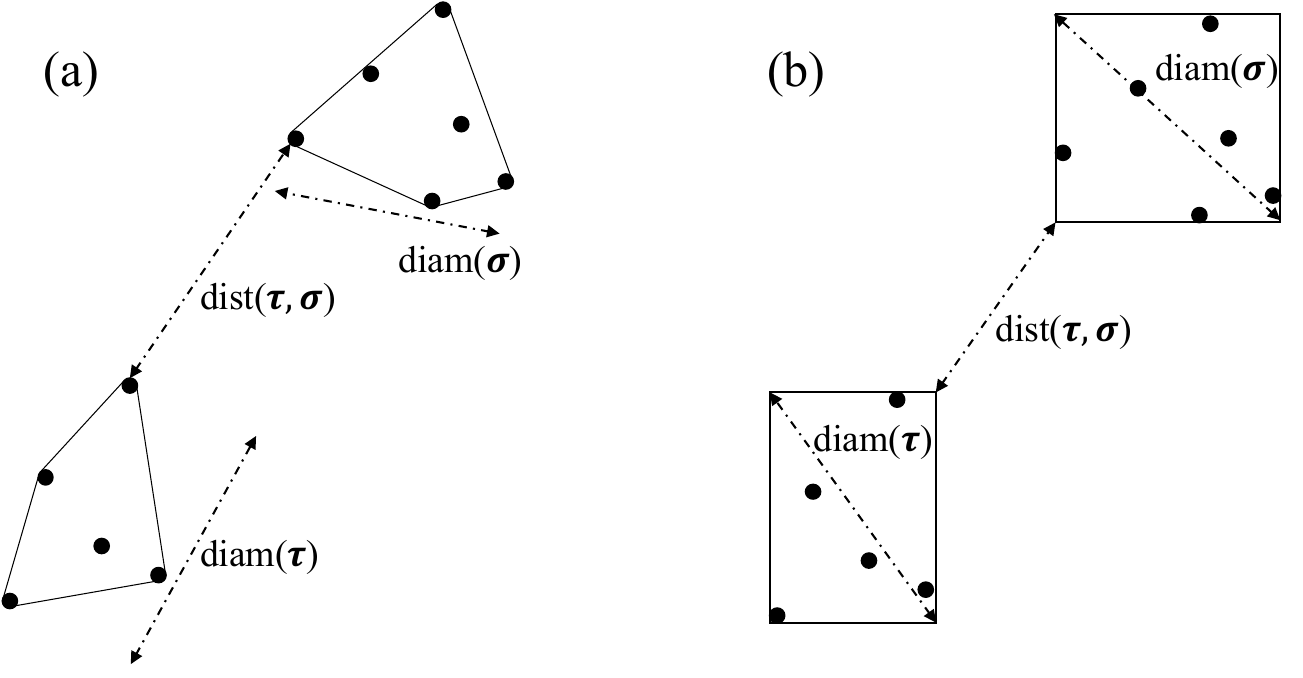}
    \caption{The diameter and distance of two clusters: (a) definition, (b) approximation by bounding box.}
    \label{fig:diam_dist}
\end{figure}

Using a cluster tree and the admissibility condition, we generate two sets of blocks: admissible blocks ($\mathcal{P}^G$) and inadmissible blocks ($\mathcal{P}^H$). These sets are generated following a recursive procedure, as outlined in Algorithm \ref{alg:Partition}.
\begin{algorithm}
\caption{BuildPartition$(\boldsymbol \tau, \boldsymbol \sigma)$}\label{alg:Partition}
\begin{algorithmic}
\If{$(\boldsymbol \tau, \boldsymbol \sigma)$ is admissible} 
    \State $S(\boldsymbol \tau, \boldsymbol \sigma) = \emptyset $
    \State Add $(\boldsymbol \tau, \boldsymbol \sigma)$ into $\mathcal{P}^G$
\ElsIf{$|\boldsymbol \tau| < C_\text{leaf}$ \textbf{and} $|\boldsymbol \sigma| < C_\text{leaf}$} 
    \State $S(\boldsymbol \tau, \boldsymbol \sigma) = \emptyset $
    \State Add $(\boldsymbol \tau, \boldsymbol \sigma)$ into $\mathcal{P}^H$
\Else
    \If{$|\boldsymbol \sigma| < C_\text{leaf}$}
    \State $S(\boldsymbol \tau, \boldsymbol \sigma) = \{ (\boldsymbol \tau', \boldsymbol \sigma) : \boldsymbol \tau' \in \text{Sons}(\boldsymbol \tau) \}$
    \ElsIf{$|\boldsymbol \tau| < C_\text{leaf}$}
    \State $S(\boldsymbol \tau, \boldsymbol \sigma) = \{ (\boldsymbol \tau, \boldsymbol \sigma') : \boldsymbol \sigma' \in \text{Sons}(\boldsymbol \sigma) \}$
    \Else
    \State $S(\boldsymbol \tau, \boldsymbol \sigma) = \{ (\boldsymbol \tau', \boldsymbol \sigma') : \boldsymbol \tau' \in \text{Sons}(\boldsymbol \tau), \boldsymbol \sigma' \in \text{Sons}(\boldsymbol \sigma) \}$
    \EndIf
    \For{$(\boldsymbol \tau', \boldsymbol \sigma') \in S(\boldsymbol \tau, \boldsymbol \sigma) $}
        \State BuildPartition$(\boldsymbol \tau', \boldsymbol \sigma')$
    \EndFor
\EndIf
\end{algorithmic}
\end{algorithm}
The process begins at the root of the cluster tree, where $\boldsymbol \tau = \boldsymbol \sigma = \mathbf{I}$. For each candidate block $(\boldsymbol \tau, \boldsymbol \sigma)$, the admissibility condition is first evaluated. If the condition is met, the block is added to the set of admissible blocks, $\mathcal{P}^G$. Otherwise, the algorithm recursively examines all sons of both $\boldsymbol \tau$ and $\boldsymbol \sigma$ until all leaves are reached, i.e. $|\boldsymbol \tau| < C_\text{leaf}$ and $|\boldsymbol \sigma| < C_\text{leaf}$. Any leaves that do not satisfy the admissibility condition are added into the set of inadmissible blocks, $\mathcal{P}^H$. Finally, the complete sets of $\mathcal{P}^G$ and $\mathcal{P}^H$ are obtained. Figure \ref{fig:1d_example} illustrates this process using a one-dimensional example with the root $\mathbf{I} = [0, 1]$ and the admissibility parameter $\eta = 1$.
\begin{figure}
\centering
\begin{subfigure}{1.0\textwidth}
\centering
\includegraphics[width=0.6\textwidth]{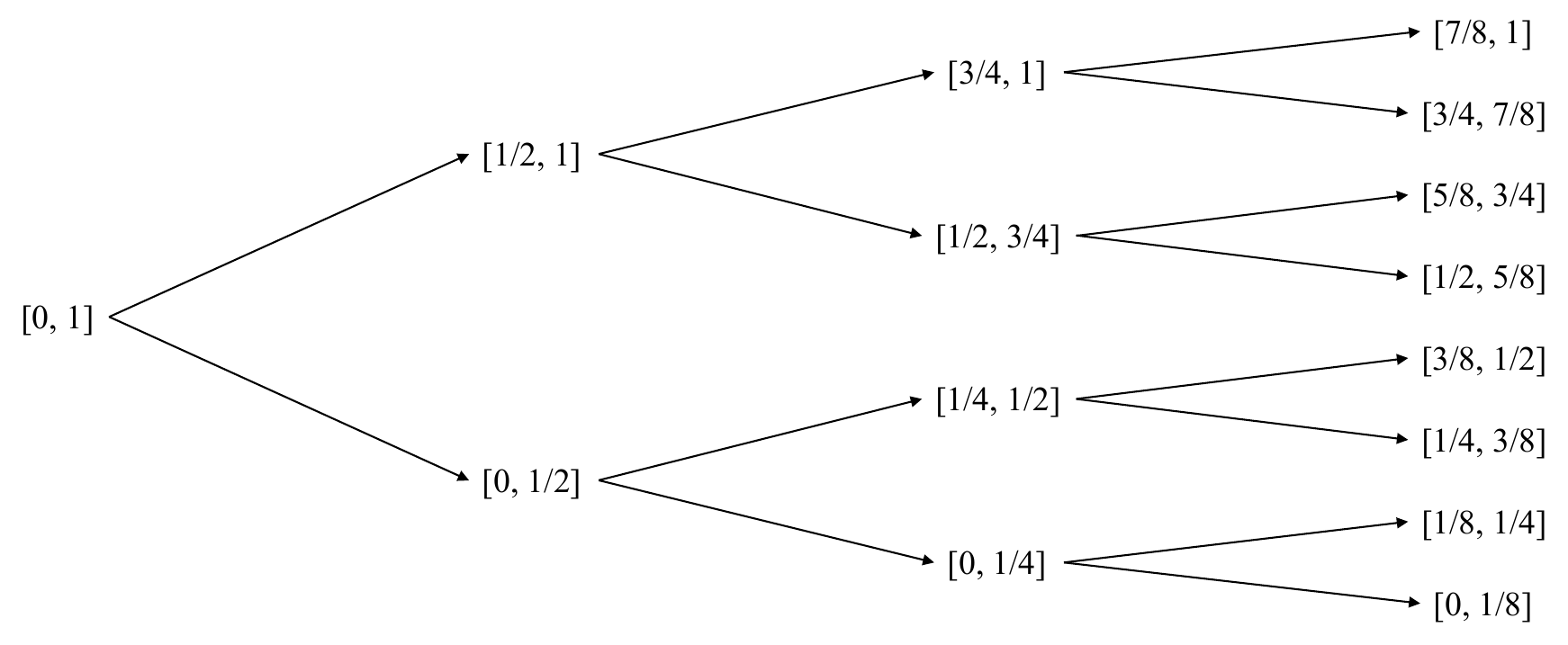}
\caption{Cluster tree built for an one-dimensional case with root $\mathbf{I} = [0, 1]$.}
\label{sfig:cluster_tree_1d}
\end{subfigure}

\vspace{0.8cm}
\begin{subfigure}{1.0\textwidth}
\centering
\includegraphics[width=0.75\textwidth]{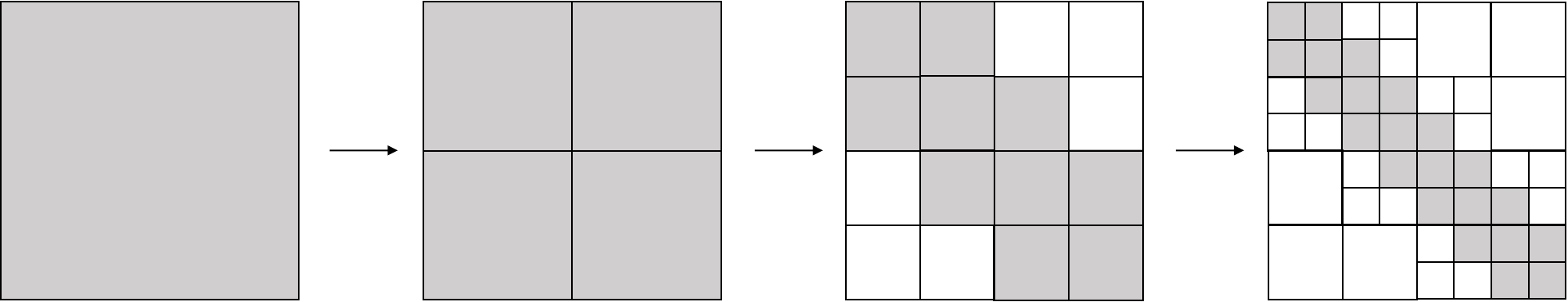}
\caption{Block partitioning corresponding to the above cluster tree and the admissibility condition in Eq.~\eqref{Eq:admissibility} with $\eta = 1$. Here, the white and gray blocks represent the admissible and inadmissible blocks, respectively.}
\label{sfig:matrix_partition_1d}
\end{subfigure}
\caption{Example of cluster tree and block partitioning with admissible and inadmissible blocks.}
\label{fig:1d_example}
\end{figure}

Note that $\boldsymbol \alpha_2^{(t)} (\mathbf{X}_i, \mathbf{X}_j)$ is essentially a function of $\frac{1}{|\mathbf{X}_j-\mathbf{X}_i|}$ \cite{Mazur1982ManysphereHI}, which has been proven asymptotically smooth \cite{Tyrtyshnikov1996cross_proof1}. Thus, if a block ($\boldsymbol \tau $, $ \boldsymbol \sigma$) is admissible, it can be approximated by a low-rank matrix \cite{Bebendorf2008Hmatrix, Tyrtyshnikov1996cross_proof1, Tyrtyshnikov2000cross_proof2}. Throughout this work, $\eta$ in Eq. \eqref{Eq:admissibility} is fixed at 1.0 consistently.

\subsection{Adaptive cross approximation (ACA)}
Following the identification of admissible blocks, the subsequent task is to compute a low-rank approximation for each block. This is accomplished through the application of the ACA method \cite{Tyrtyshnikov2000cross_proof2}. The basic idea of ACA is to approximate a matrix with a product of one column and one row of the matrix and then repeat this process until the approximation with desired precision is achieved.

Considering an admissible block $\mathbf{G} \in  \mathbb{R}^{c \times q} $, the goal of ACA is to approximate $\mathbf{G} \in  \mathbb{R}^{c \times q} $ by a sequence of rank-1 products as:
\begin{equation}
    \mathbf{G} \approx \mathbf{C} \mathbf{Q} = \sum\limits_{k = 1}^{l}  \mathbf{c}_k \mathbf{q}^T_k \;,
\end{equation}
where $\mathbf{c}_k  \in  \mathbb{R}^{c \times 1} $ and $\mathbf{q}^T_k  \in  \mathbb{R}^{1 \times q} $ are column and row vectors, respectively. At each iteration $k$, we select a column $\mathbf{c}_k$ (pivot $j_k$ from previously unselected columns) and a row $\mathbf{q}^T_k$ (pivot $i_k$ from previously unselected rows) from the residual matrix $\mathbf{E}_{k-1} =  \mathbf{G} - \sum_{i = 1}^{k- 1} \mathbf{c}_i \mathbf{q}_i^T$. ACA determines $j_k$ and $i_k$ by only examining the previously identified row and column, respectively, which effectively avoids computing the entire residual matrix. Specifically, at each iteration $k$, the row pivot $i_k$ used in the current iteration and the column pivot $j_{k + 1}$ for the next iteration are determined by:
\begin{equation}\label{Eq:pivot_rule}
\begin{split}
    i_k & = \argmax_{i \neq i_1, \dots, i_{k - 1}} | \mathbf{E}_{k - 1}(i, j_k)| \\
    j_{k + 1} & = \argmax_{j \neq j_1, \dots, j_{k}} |\mathbf{E}_{k - 1}(i_k, j)| 
\end{split}
\end{equation}
where $| \cdot |$ denotes the determinant of a matrix. Notably, unlike standard $\mathcal{H}$-matrix definitions where elements are scalars, each element in our matrix is a $3 \times 3$ submatrix, as previously explained. In ACA, the iteration process starts with a randomly selected initial column index, $j_1$, and terminates when the magnitude of a new update is sufficiently small, i.e.,
\begin{equation}
    \nu \leq \zeta \lambda\;, \quad \text{with}~~\nu = \| \mathbf{c}_k \mathbf{q}^T_k \|_F  \approx \| \mathbf{G} - \mathbf{C}\mathbf{Q} \|_F\;, ~~ \lambda = \|\mathbf{C}\mathbf{Q} \|_F \approx \| \mathbf{G} \|_F  \;,
    \label{equ:criterion_ACA}
\end{equation}
where $\| \cdot \|_F$ denotes the Frobenius norm; $\zeta$ is the preset tolerance. The complete procedure of ACA employed in this work is presented in Algorithm \ref{alg:ACA}. 
Some notations used therein are explained below. Rows and columns of a matrix $\mathbf{A}$ are denoted as $\mathbf{A}(I, :) $ and $\mathbf{A}(:, I) $, respectively, where $I$ is the index set. Vertical and horizontal concatenations of $\mathbf{A}$ and $\mathbf{B}$ are represented by $[\mathbf{A}; \mathbf{B}]$ and $[\mathbf{A}, \mathbf{B}]$, respectively. As the $i$-th element of $\mathbf{c}$ is denoted by $\mathbf{c}(i)$ (a $3\times 3$ submatrix), the inverse of $\mathbf{c}(i)$ is written as  $[\mathbf{c}(i)]^{-1}$.  
\begin{algorithm}
\caption{ACA}\label{alg:ACA}
\begin{algorithmic}[1]
\Require Subroutine to calculate entries of $\mathbf{G} \in \mathbb{R}^{c \times q} $, tolerance $\epsilon$ 
\Ensure Low-rank approximation of $\mathbf{G} \approx \mathbf{C}\mathbf{Q}$ with rank $l$
\State $\mathbf{C} = \emptyset$, $\mathbf{Q} = \emptyset$, $\lambda = 0$
\State Randomly select the initial column index $j_1$ from $\{1, 2, \dots, q \}$
\For{$k = 1$ to $\min\{c, q\}$}
\State $\mathbf{c}_k = \mathbf{E}_{k - 1}(:, j_k) = \mathbf{G}(:, j_k) - \mathbf{C}\mathbf{Q}(:, j_k) $
\State $i_k = \underset{i \neq i_1, \dots, i_{k - 1}}\argmax \|\mathbf{c}_k(i) \|_F$
\State $\mathbf{c}_k = \mathbf{c}_k \cdot [\mathbf{c}_k(i_k)]^{-1} $
\State $\mathbf{q}^T_k = \mathbf{E}_{k - 1}(i_k,:) = \mathbf{G}(i_k,:) - \mathbf{C}(i_k,:)\mathbf{Q} $
\State $j_{k + 1} = \underset{j \neq j_1, \dots, j_{k}}\argmax \|\mathbf{q}_k(j)\|_F$
\State $\nu^2 = \| \mathbf{c}_k \mathbf{q}^T_k \|^2_F$
\State $\lambda^2 = \lambda^2 + \nu^2 + 2 \sum_{i = 1}^{k - 1} \mathbf{Q}(i,:) \mathbf{q}_k \mathbf{c}^T_k \mathbf{C}(:,i)$
\State $\mathbf{C} = [\mathbf{C}, \mathbf{c}_k]$, $\mathbf{Q} = [\mathbf{Q}; \mathbf{q}^T_k]$. 
\State Terminate if $\nu < \zeta \lambda$
\EndFor
\end{algorithmic}
\end{algorithm}

From Algorithm \ref{alg:ACA}, it is evident that an $l$-rank approximation can be constructed by evaluating only the entries on the $l$ pivoting rows and columns, thereby eliminating the need to examine all entries of $\mathbf{G}$. As a result, the computational complexity of ACA scales as $O((c+q)l^2)$.

\subsection{Parallel implementation}
Given the GPU-centric nature of our calculations (neural network inference and matrix-vector multiplication), this section elaborates on the GPU programming model, the strategy for addressing GPU latency, the approach for workload distribution across GPUs, and inter-GPU communication. 

First of all, the Kokkos \cite{9485033} programming model is employed to harness the high flexibility of parallel execution capabilities offered by GPUs. More specifically, \textbf{View} is used for storing data, and \textbf{parallel\_for} is employed to manage parallel operations on GPUs. A two-level parallelism model is adopted in our implementation: the outer level spans over the blocks, while the inner level deals with the elements within a block. The number of iteration steps in the inner level is not necessarily the same across different blocks. The alignment of the block element index to the thread index of hardware is automatically taken care by Kokkos. For certain inner parallel level operations, such as summations within a block and maximum argument determinations required by the ACA algorithm, \textbf{parallel\_reduce} is used. 

Next, after $\mathbf{M}^{(t, 2)}$ is partitioned into $N^G$ admissible blocks ($\mathbf{G}^{(\bm{\tau}_i^G, \bm{\sigma}_i^G)}$) and $N^H$ inadmissible blocks ($\mathbf{H}^{(\bm{\tau}_i^H, \bm{\sigma}_i^H)}$), careful coordination of entry evaluation and matrix-vector multiplication is required to address GPU latency. The inherent high latency of GPUs implies that the evaluation of entries of $\mathbf{G}^{(\bm{\tau}_i^G, \bm{\sigma}_i^G)}$ and $\mathbf{H}^{(\bm{\tau}_i^H, \bm{\sigma}_i^H)}$ cannot be performed sequentially, and multiple blocks have to be evaluated in parallel to take the advantage of the high memory bandwidth provided by GPUs. Due to the extensive number of queries required for evaluating the entries of $\mathbf{G}$ and $\mathbf{H}$, these queries are coordinated into multiple iterations. In each iteration, the entries of a selected group of blocks are evaluated by querying the HIGNN; the corresponding velocity vector $\mathbf{U}^{\bm{\tau}_i}$ is then updated accordingly. To optimally utilize the GPUs' memory, the max number of queries in each iteration is limited by $N_{\max}^{\text{query}}$ which is determined according to the memory size on the GPU and can vary on different platforms. The memory allocated for query operations, including the intermediate storage for $\mathbf{C}$ and $\mathbf{Q}$ used by ACA, scales linearly with $N_{\max}^{\text{query}}$, and can be held constant as the number of particles $N$ increases. Therefore, with the exception of memory allocated for the velocity and force vectors $\mathbf{U}$ and $\mathbf{F}$, the implementation's memory footprint remains constant and does not scale with $N$. The proposed implementation approach is portable to most of the currently available GPU-equipped platforms, regardless of GPU count. The specific workload distributions across GPUs for inadmissible and admissible blocks, respectively, are discussed as follows.

\subsubsection{Inadmissible blocks}
 
The workload of inadmissible blocks is evenly distributed among $N_{\text{GPU}}$ GPUs. The total workload can be measured with respect to the sizes of inadmissible blocks as $W^H = \sum_{i=1}^{N^H} |\bm{\tau}_i^H| \times |\bm{\sigma}_i^H|$, where $|\bm{\sigma}^H_i|$ and $|\bm{\tau}^H_i|$ denote the number of particles in set (cluster) $\bm{\sigma}^H_i$ and $\bm{\tau}^H_i$, respectively. The resultant partition of blocks on the $g$-th GPU is then denoted as $\mathcal{P}^H_g = \{ (\bm{\tau}^H_1, \bm{\sigma}^H_1), (\bm{\tau}^H_2, \bm{\sigma}^H_2), \cdots (\bm{\tau}^H_{|\mathcal{P}_g^H|}, \bm{\sigma}^H_{|\mathcal{P}_g^H|}) \}$, and the corresponding workload is $W_g^H = \sum_{i=1}^{|\mathcal{P}_g^H|} |\bm{\tau}_i^H| \times |\bm{\sigma}_i^H| \approx \frac1{N_\text{GPU}} W^H$.

Typically, a single GPU cannot manage the entire $W_g^H$ workload in a single iteration. Thus, it is necessary to further partition the workload $W_g^H$ into several groups and process them sequentially. In each iteration, a sub group $\widetilde{\mathcal{P}}^H_g$ is selected from $\mathcal{P}_g^H$ such that each $(\bm{\tau}_i^H, \bm{\sigma}_i^H) \in \widetilde{\mathcal{P}}^H_g$ has not yet been evaluated. These chosen blocks are aggregated to optimally fill the capacity $N^{\text{query}}_{\max}$, ensuring the workload of the specific iteration $\widetilde{W}^H_g = \sum_{i = i_1}^{i_2} |\bm{\tau}_i^H| \times |\bm{\sigma}_i^H|$ is close to $N^{\text{query}}_{\max}$. As querying the GNN for $\bm{\alpha}^{(t)}_2$ requires two coordinates as arguments, each entry in matrix $\mathbf{H}^{(\bm{\tau}_i^H, \bm{\sigma}_i^H)}_{mn}$ is re-indexed as $k = \sum_{j = i_1}^{i-1} |\bm{\tau}_j^H| \times |\bm{\sigma}_j^H| + p \times |\bm{\tau}_i^H| + n$, such that $\mathbf{y}_k = \mathbf{X}^{\bm{\tau}_i^H}_p$ and $\mathbf{z}_k = \mathbf{X}^{\bm{\sigma}_i^H}_n$. These two coordinates are then collected into the relative coordinate matrix $\mathbf{Y}, \mathbf{Z} \in \mathbb{R}^{\widetilde{W}_g^H \times 3}$, such that $\mathbf{Y}_k = \mathbf{y}_k^\intercal$ and $\mathbf{Z}_k = \mathbf{z}_k^\intercal$. By querying the GNN, we obtain each entry of $\mathbf{H}^{(\bm{\tau}_i^H, \bm{\sigma}_i^H)}$, which contains $|\bm{\tau}_i^H| \times |\bm{\sigma}_i^H|$ elements in total, with each element a $3 \times 3$ sub-matrix following the definition of $\mathbf{M}^{(t, 2)}$. Finally, the matrix-vector multiplication for $\mathbf{U}^{\bm{\tau}^H_i} = \mathbf{H}^{(\bm{\tau}^H_i, \bm{\sigma}^H_i)} \cdot \mathbf{F}^{\bm{\sigma}^H_i}$ is performed for each block $(\bm{\tau}_i^H, \bm{\sigma}_i^H) \in \mathcal{P}^H$ simultaneously.

\subsubsection{Admissible blocks}

The workload of admissible blocks is measured as $W^G = \sum_{i=1}^{N^G} \max(| \bm{\tau}_i^G |, | \bm{\sigma}_i^G |)$, noting that the ACA algorithm only selects one row and one column in each iteration, and the number of iterations required to meet the termination criterion is independent of the block dimensions. The total workload is evenly distributed among $N_\text{GPU}$ GPUs. Denote $\mathcal{P}_g^G = \{ (\bm{\tau}_1^G, \bm{\sigma}_1^G), (\bm{\tau}_2^G, \bm{\sigma}_2^G), \cdots (\bm{\tau}^G_{|\mathcal{P}_g^G|}, \bm{\sigma}^G_{|\mathcal{P}_g^G|}) \}$ as the set of admissible blocks on the $g$-th GPU. The workload on the $g$-th GPU can then be expressed as: $W_g^G = \sum_{i=1}^ {|\mathcal{P}_g^G|} \max(| \bm{\tau}_i^G |, | \bm{\sigma}_i^G |) \approx \frac1{N_\text{GPU}} W^G$. 

Apart from the memory required for storing relative coordinates and GNN query results, the ACA algorithm (Algorithm \ref{alg:ACA}) necessitates additional space for intermediate results $\mathbf{C}$ and $\mathbf{Q}$. To pre-allocate this memory and prevent dynamic allocation during execution, two spare rank-2 \textbf{View}s, $\mathcal{C}$ and $\mathcal{Q}$, are created. Each \textbf{View} has dimensions $(N_{\max}^{\text{ite}} \times N^{\text{query}}_{\max}, 9)$, where $N_{\max}^{\text{ite}}$ represents the maximum allowed iterations for the ACA algorithm, and each $3 \times 3$ submatrix within $\mathbf{C}$ and $\mathbf{Q}$ is flattened into a $1 \times 9$ vector according to the right layout rule. Based on our numerical experiments, the number of iterations in the ACA algorithm is highly correlated with the tolerance $\zeta$, remaining nearly constant across all admissible blocks and invariant to different particle configurations. Therefore, $N_{\max}^{\text{ite}}$ can be empirically determined once $\zeta$ is pre-set.

To process all $|\mathcal{P}_g^G|$ blocks for low-rank approximation and matrix-vector multiplication, a two-level loop is employed. The outer loop iterates through the blocks, selecting a subgroup $\widetilde{\mathcal{P}}_g^G$ from the partition $\mathcal{P}^G_g$ for evaluation in each iteration. The workload of these selected blocks, estimated as $\widetilde{W}_g^G = \sum_{i \in \widetilde{\mathcal{P}}_g^G} \max(|\bm{\tau}_i^G|, |\bm{\sigma}_i^G|)$, is chosen to approximate $N_{\max}^{\text{query}}$, ensuring that the storage for intermediate results $\mathbf{C}$ and $\mathbf{Q}$ does not exceed the allocated size of the two \textbf{View}s. The inner loop then applies the ACA algorithm, detailed in the next paragraph, to each block in parallel. Low-rank approximations are computed concurrently, with intermediate data stored in the two \textbf{View}s. Following the inner loop, the matrix-vector multiplication $\mathbf{U}^{\bm{\tau}^G_i} = \mathbf{C}^{(\bm{\tau}_i^G, \bm{\sigma}^G_i)} \cdot \mathbf{Q}^{(\bm{\tau}^G_i, \bm{\sigma}^G_i)} \cdot \mathbf{F}^{\bm{\sigma}^G_i}$ is performed. The two \textbf{View}s are subsequently cleared for use in the next outer iteration.

Within the inner loop, the ACA algorithm (Algorithm \ref{alg:ACA}) is parallelized on GPUs. Lines 2-12 are encapsulated into separate parallel code blocks. Each line, except 4 and 7, corresponds to a \textbf{parallel\_for} operation. For operations involving element-wise manipulation, such as line 5, a nested parallelism model is used: the outer loop iterates over selected blocks, while the inner loop manages elements within each block using GPU threads. Lines 4 and 7, which require accessing the GNN for $\bm{\alpha}_2^{(t)}$, are executed in three stages. First, relative coordinates for blocks in $\widetilde{\mathcal{P}}_g^G$ are computed in parallel and stored in coordinate matrices $\mathbf{Y}$ and $\mathbf{Z}$. Second, these coordinates are passed to the GNN for inference. Finally, the inference result $\bm{\alpha}_2^{(t)}(\mathbf{Y}, \mathbf{Z})$ is stored into \textbf{View}s: $\mathcal{C}$ for line 4 and $\mathcal{Q}$ for line 7.

\subsubsection{Inter-GPU communication}
Inter-GPU communication occurs only at the beginning and end of each time step. Specifically, at the beginning, the current particle configuration and applied force $\mathbf{F}$ are broadcast to all GPUs. Each GPU then performs matrix-vector multiplications on its assigned blocks, processing workloads $W_g^H$ and $W_g^G$ for inadmissible and admissible blocks, respectively. At the end of the time step, a global reduction (summation) is performed to obtain the final result $\mathbf{U}$, constituting the other instance of inter-GPU communication. These inter-GPU communication costs scale as $\mathcal{O}(N \log N_\text{GPU})$, where $N$ is the number of particles, and $N_\text{GPU}$ is the number of GPUs. Notably, both neural network inference and low-rank approximation via ACA require no inter-GPU communication. 

\section{Results}\label{sec:results}
Building upon the thoroughly validated accuracy and efficiency of HIGNN in our prior work \cite{Ma2022HIGNN,ma2024shape}, we introduce $\mathcal{H}$-matrix integration in this work to further improve it, especially its scalability for simulating large-scale systems. This enhanced framework is named as $\mathcal{H}$-HIGNN. This section provides a comprehensive evaluation of $\mathcal{H}$-HIGNN, with respect to the accuracy, computational complexity, and scalability, as detailed below.

\subsection{Assessment of accuracy}\label{subsec:acc_test}
The accuracy of the $\mathcal{H}$-HIGNN model is essentially determined by the error introduced by $\mathcal{H}$-matrix to $\mathbf{M}^{(t,2)}$. Thus, we define the following relative error:
\begin{equation}\label{Eq:error_equation}
    \epsilon_{r} = \dfrac{\| (\mathbf{M}^{(t,2)} - \mathbf{M}_{\mathcal{H}}^{(t,2)}) \mathbf{F} \|_2}{\| \mathbf{M}^{(t,2)} \mathbf{F} \|_2} = \dfrac{\| \mathbf{M}^{(t,2)} \mathbf{F} - \mathbf{M}_{\mathcal{H}}^{(t,2)} \mathbf{F} \|_2}{\| \mathbf{M}^{(t,2)} \mathbf{F} \|_2} \;,
\end{equation}
where $\mathbf{M}_{\mathcal{H}}^{(t,2)}$ stands for the $\mathcal{H}$-matrix constructed for $\mathbf{M}^{(t,2)}$, following the methodology described in \S\ref{sec:H-matrix}. Here, the error is evaluated by multiplying the matrix with a force vector $\mathbf{F}$, due to two reasons: 1) only the multiplication is needed in predicting each particle's velocity, as seen in Eq. \eqref{Eq:theo_basis_mat}; 2) instead of constructing the full matrix before multiplication, $\mathcal{H}$-matrix construction is designed to perform matrix-vector multiplication concurrently. In particular, $\mathbf{F}$ was generated randomly when calculating the error in Eq. \eqref{Eq:error_equation}. After five repetitions, the average error was used as the accuracy metric.

In this effort, particles were initially positioned to form a primitive cubic lattice \cite{birkhoff1940lattice} within an unbounded domain, with subsequent introduction of small random perturbations to each particle's position to approximate realistic conditions. The numerical experiments consider the numbers of particles ranging from 8,000 to 1,000,000, allowing for a comprehensive assessment of the model's accuracy. For error analysis, we constrain our studies to 1,000,000 particles, as we require the original HIGNN model with $O(N^2)$ computational complexity to serve as a baseline for calculating the error according to Eq.~\eqref{Eq:error_equation}. Here, two different particle distribution scenarios are considered for assessing the model's accuracy. In the first scenario, the particles' volume fractions or concentrations in suspensions are maintained constant, by preserving the interparticle spacing while increasing the number of particles in the lattice structure. This constant-concentration configuration allows us to examine the model's accuracy for systems of varying sizes under consistent spatial distributions of particles. In the second scenario, we initially generated the cubic lattice comprising 1,000,000 particles, from which subsets of particles were randomly selected. This way we can investigate the effect of varying particle concentrations on the model's accuracy. The results corresponding to these two scenarios are presented in Figures~\ref{fig:error_zeta} and \ref{fig:comple_Cleaf}, where the effects of varying $\zeta$, the tolerance for ACA in Eq. \eqref{equ:criterion_ACA}, and $C_\text{leaf}$, the leaf node size, are also examined. 

It can be seen that the value of $\zeta$ significantly influences the model's accuracy, as it directly controls the accuracy of the low-rank approximation for each admissible block. Given that non-admissible blocks exhibit zero error, the overall relative error $\epsilon_{r}$, which can be interpreted as a weighted average of errors across both admissible and non-admissible blocks, typically remains significantly lower than the value of $\zeta$. Consequently, $\zeta$ serves as an effective upper bound for the overall relative error of $\mathcal{H}$-HIGNN. On the other hand, despite increased partitioning of $\mathbf{M}^{(t,2)}$ into smaller blocks by reducing $C_\text{leaf}$, which dictates the maximum particles per leaf, the error increase is minimal. This robustness, stemming from a slight increase in admissible blocks,  demonstrates the $\mathcal{H}$-matrix's resilience to finer block partitioning. 

Figures \ref{sfig:error_zeta_fixed_density} and \ref{sfig:error_Cleaf_fixed_density} show elevated errors between $10^5$ and $10^6$ particles. This behavior is due to the growing problem domain when $N$ increases while maintaining a constant interparticle distance. The growing domain alters the cluster tree construction, as the spatial extent represented by each node on the tree varies with $N$. This variation directly affects the partitioning of admissible blocks. Specifically, the elevated errors coincide with instances where the cluster tree produces a greater number of admissible blocks, a point we will further validate through complexity testing in the next section. In contrast, when the problem domain is held constant and particle concentration increases with $N$, i.e., the second scenario of particle distributions, the error curves are significantly flatter, as depicted in Figures \ref{sfig:error_zeta_fixed_domain} and \ref{sfig:error_Cleaf_fixed_domain}. The fixed domain ensures a more consistent cluster tree structure across different $N$ values, as the spatial extent of most nodes remains relatively stable. This consistency translates to a stable proportion of admissible blocks, resulting in a more uniform error distribution.

\begin{figure}[htbp]
    \centering
    \begin{subfigure}{0.48\textwidth}
    \centering
    \includegraphics[width = 8.0cm]{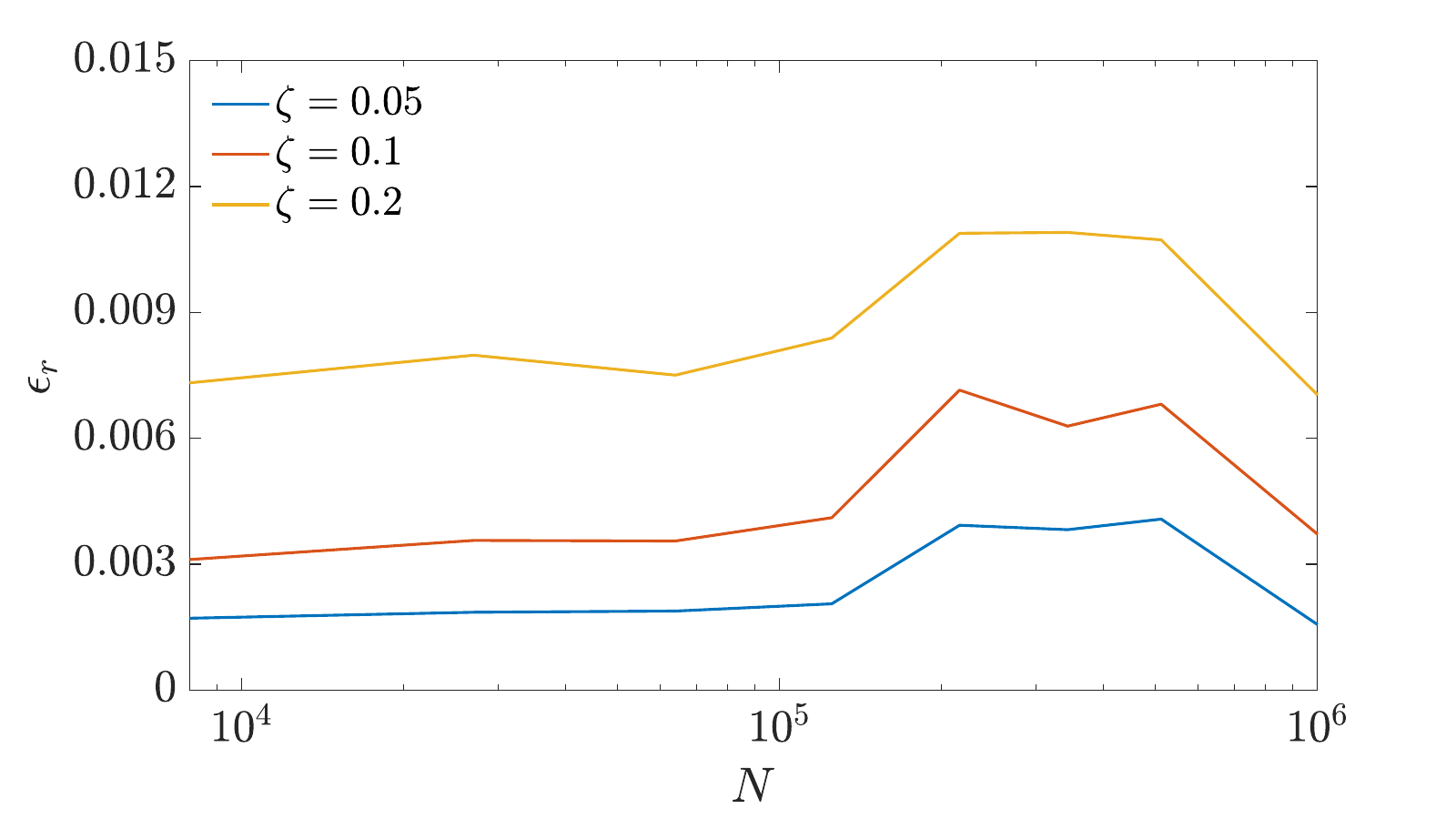}
    \caption{Fixed volume fraction with increasing domain size}
    \label{sfig:error_zeta_fixed_density}
    \end{subfigure}
    \quad
    \begin{subfigure}{0.48\textwidth}
    \centering
    \includegraphics[width = 8.0cm]{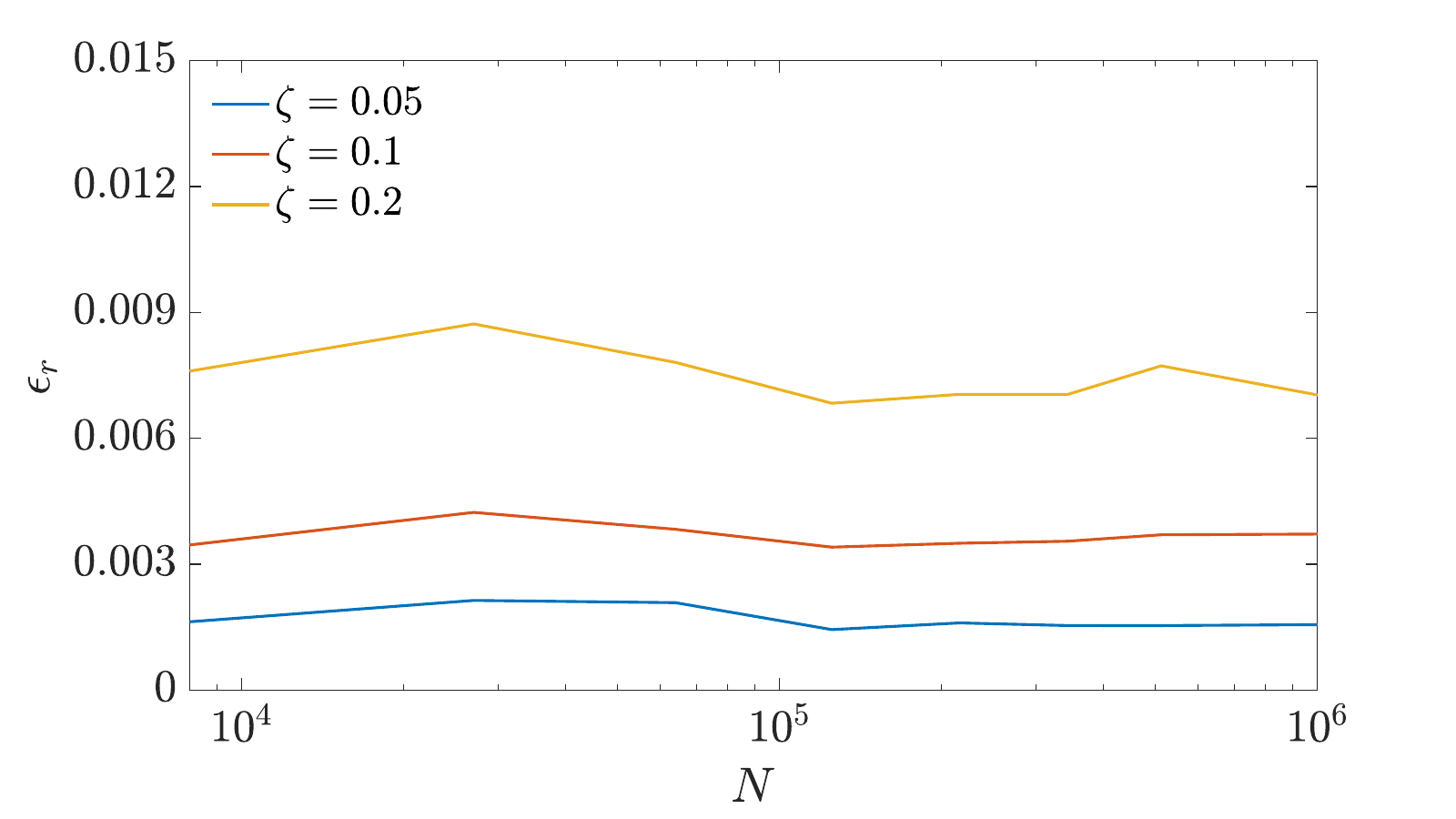}
    \caption{Fixed domain with increasing volume fraction}
    \label{sfig:error_zeta_fixed_domain}
    \end{subfigure}
    \caption{The accuracy of $\mathcal{H}$-HIGNN evaluated by the relative error defined in Eq. \eqref{Eq:error_equation} with different $\zeta$ but fixed $C_\text{leaf} = 50$.} 
    \label{fig:error_zeta}
\end{figure}

\begin{figure}[htbp]
    \centering
    \begin{subfigure}{0.48\textwidth}
    \centering
    \includegraphics[width = 8.0cm]{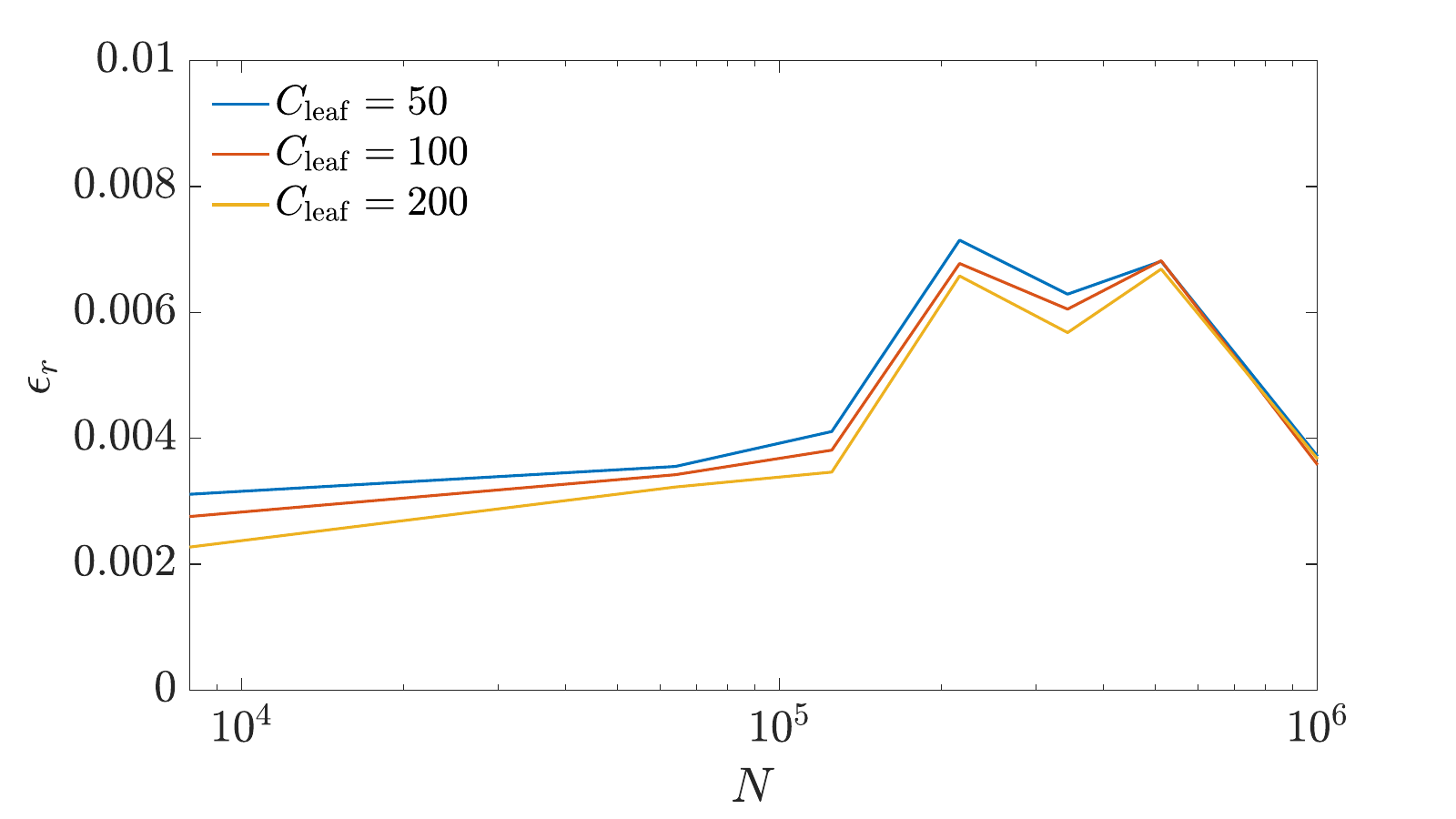}
    \caption{Fixed volume fraction with increasing domain size}
    \label{sfig:error_Cleaf_fixed_density}
    \end{subfigure}
    \quad
    \begin{subfigure}{0.48\textwidth}
    \centering
    \includegraphics[width = 8.0cm]{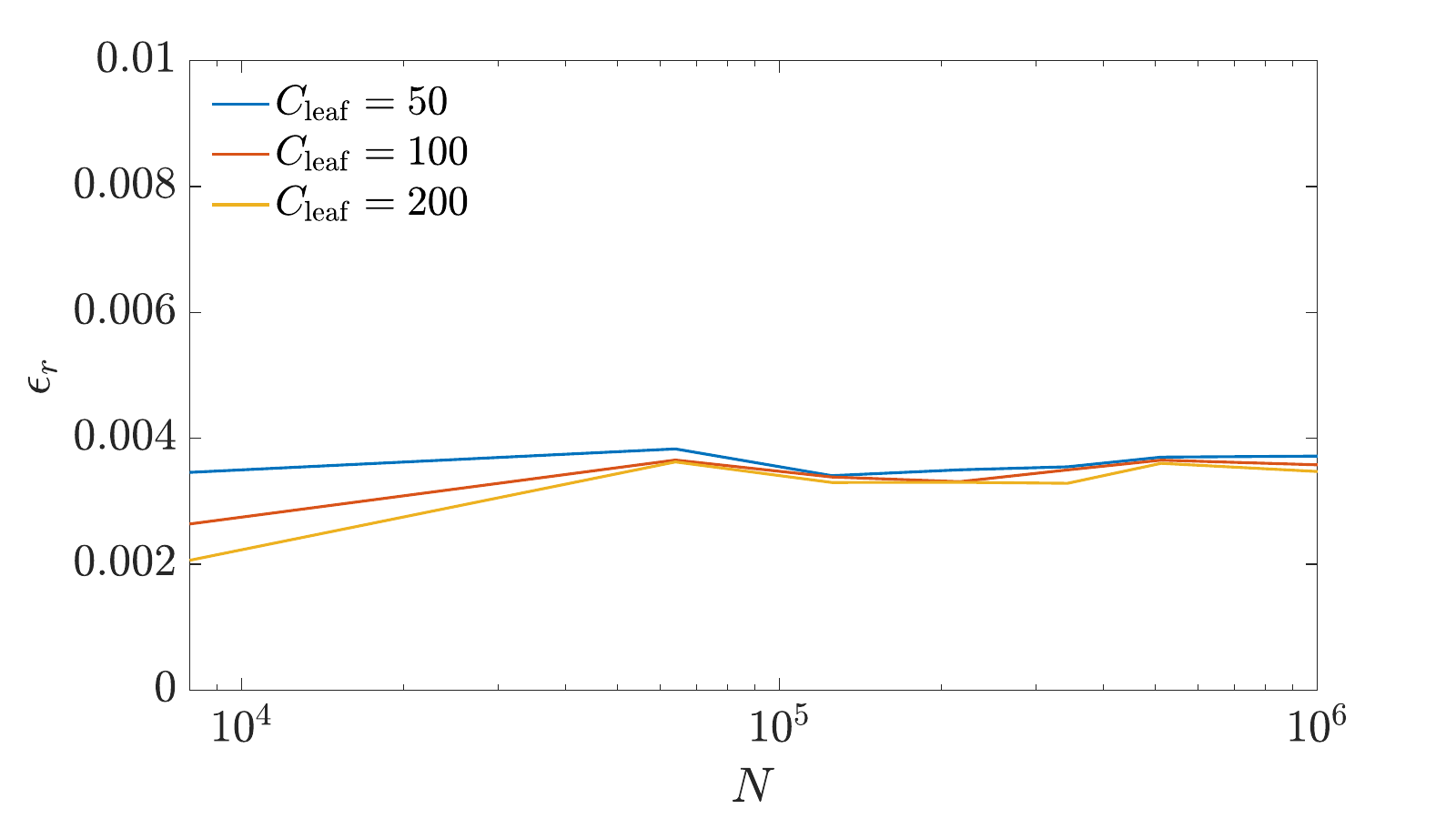}
    \caption{Fixed domain with increasing volume fraction}
    \label{sfig:error_Cleaf_fixed_domain}
    \end{subfigure}
    \caption{The accuracy of $\mathcal{H}$-HIGNN evaluated by the relative error defined in Eq. \eqref{Eq:error_equation} with different $C_\text{leaf}$ but fixed $\zeta = 0.1$.} 
    \label{fig:comple_Cleaf}
\end{figure}

\subsection{Assessment of complexity and scalability}
We next assess the computational complexity of the $\mathcal{H}$-HIGNN model. It can be evaluated by examining the total number of submatrices inferred through GNN during the construction of the $\mathcal{H}$-HIGNN model. Thus, we compare the number of queries for two-body HIs $\boldsymbol{\alpha}_2^{(t)}$ by $\mathcal{H}$-matrix with the combined sum of other HIs, i.e. $\boldsymbol{\alpha}_2^{(s)}$, $\boldsymbol{\alpha}_3^{(t)}$ and $\boldsymbol{\alpha}_3^{(s)}$, as shown in 
Figure~\ref{fig:2body_vs_3body}. We extended the system size to 10 million particles for this section's analysis, enabling a more comprehensive investigation on the complexity and scalability. The results in Figure~\ref{fig:2body_vs_3body} reveal that $\boldsymbol{\alpha}_2^{(t)}$ dominates the overall computational complexity, exhibiting 1-2 orders of magnitude more queries than the other interaction terms. And we are able to reduce its complexity to $O(N \log N)$ from $O(N^2)$ through integrating $\mathcal{H}$-matrix. In light of its dominant role, in the following tests, we focus on the complexity and scalability of $\boldsymbol{\alpha}_2^{(t)}$ under varying $\mathcal{H}$-matrix parameters. 
\begin{figure}[htbp]
    \centering
    \includegraphics[width = 9cm]{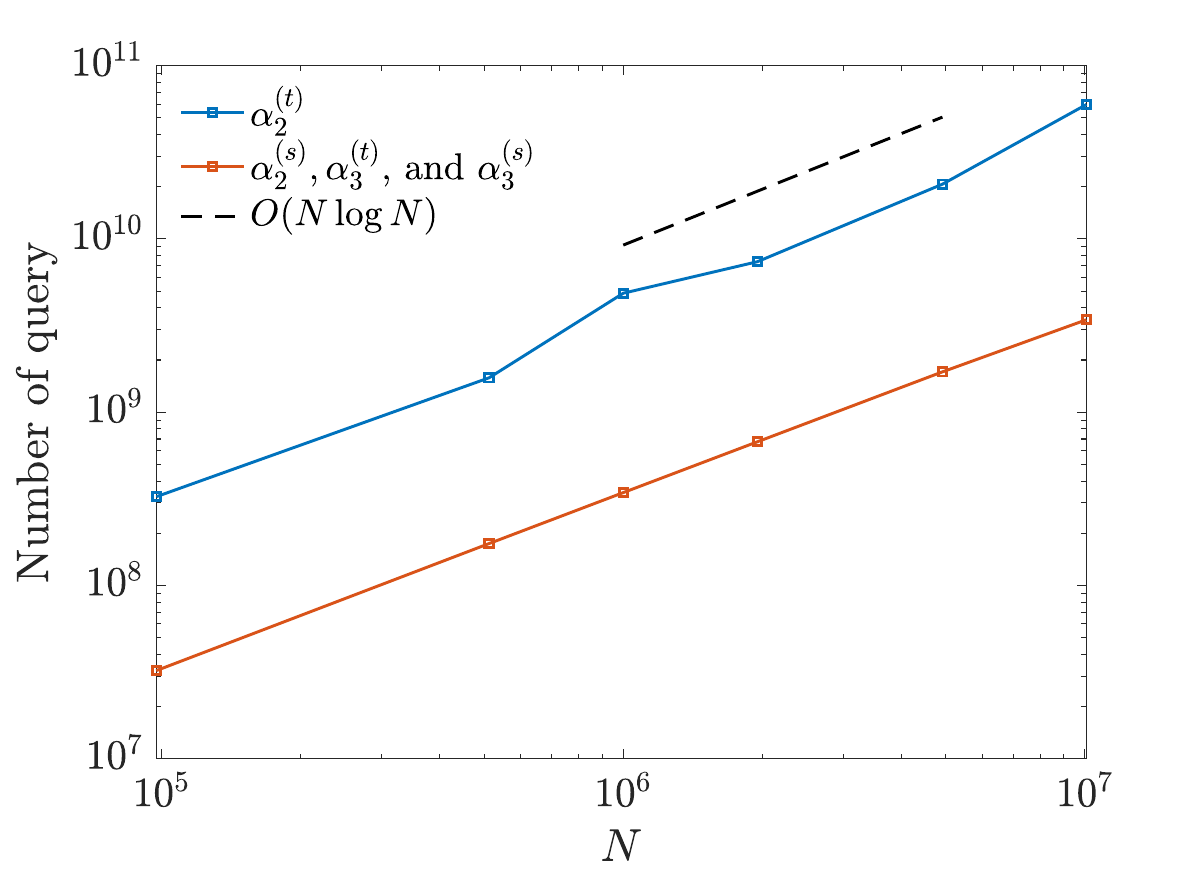}
    \caption{Number of queries of $\boldsymbol{\alpha}_2^{(t)}$ in $\mathcal{H}$-matrix compared with the number of queries of other terms during the inference stage. Here, the $\mathcal{H}$-matrix parameters are set as: $\zeta = 0.1$ and $C_\text{leaf} = 50$. Also, only the scenario of fixed volume fraction (with increasing domain size) is considered here, as the volume fraction can significantly affect the number of neighbors for 3-body HIs.}
    \label{fig:2body_vs_3body}
\end{figure}

Figures~\ref{fig:complexity_zeta} and \ref{fig:complexity_Cleaf} present the computational complexity analysis across different parameter settings. Using the same two scenarios for particle configurations as described in \S \ref{subsec:acc_test}, the results demonstrate that the $\mathcal{H}$-matrix consistently achieves a computational complexity of $O(N\log N)$, independent of parameter choices. This represents a substantial improvement over the original HIGNN model's quadratic scaling of $O(N^2)$, leading to significant enhancement in computational efficiency, particularly for large-scale systems. For systems containing 10 million ($10^7$) particles, integrating $\mathcal{H}$-matrix into HIGNN reduces the number of required queries by 3-4 orders of magnitude. Furthermore, when considering these results together with the error analysis, improving the accuracy of $\mathcal{H}$-matrix is more effectively achieved by reducing $\zeta$ rather than increasing $C_\text{leaf}$. A decrease in $\zeta$ yields better accuracy improvements while incurring a smaller computational overhead. Therefore, in practical applications, one can optimize performance by selecting smaller leaf node limits to enhance computational efficiency while utilizing $\zeta$ as the primary parameter for error control.
\begin{figure}[htbp]
    \centering
    \begin{subfigure}{0.48\textwidth}
    \centering
    \includegraphics[width = 8.0cm]{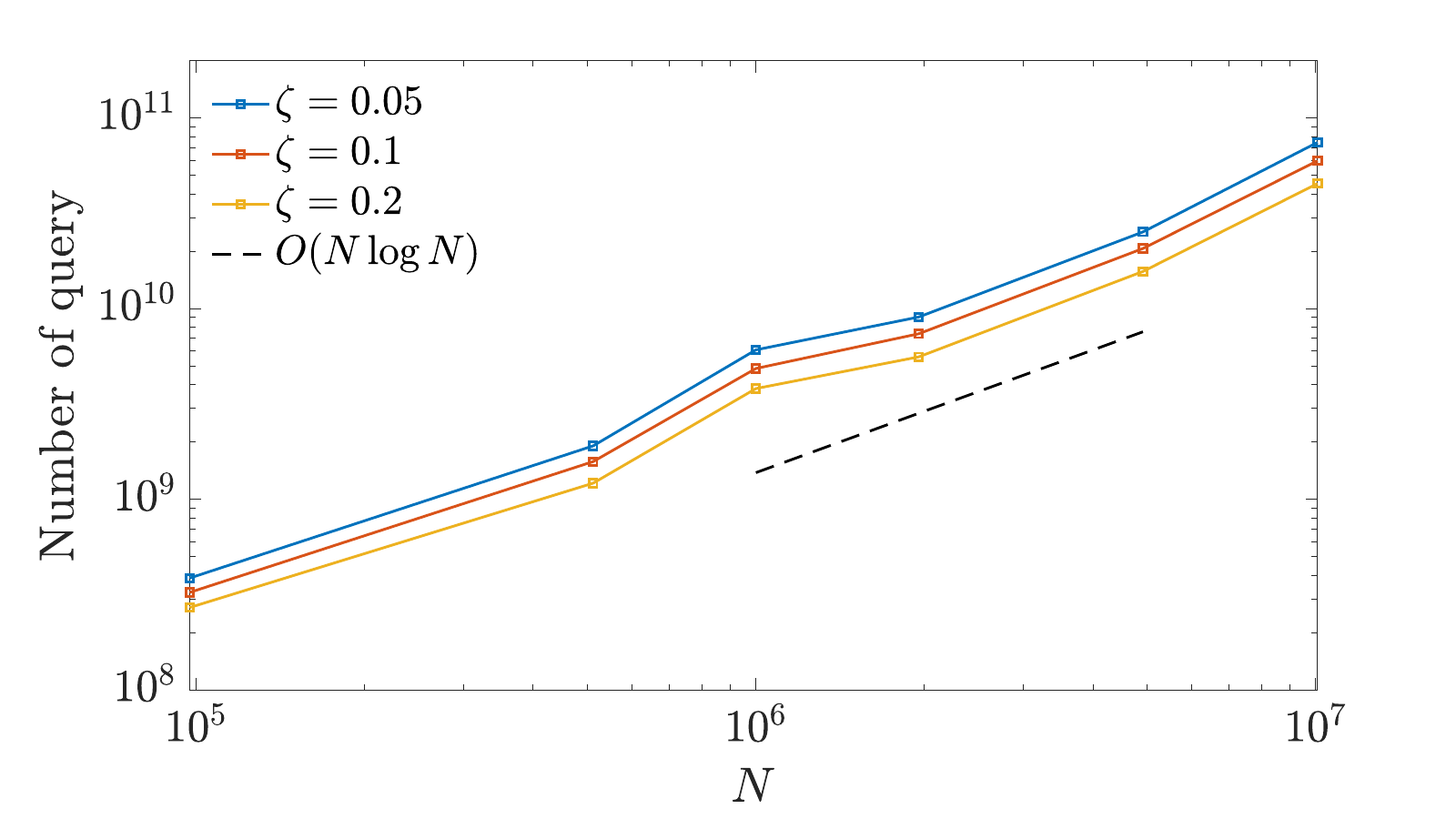}
    \caption{Fixed volume fraction with increasing domain size}
    \label{sfig:complexity_zeta_fixed_density}
    \end{subfigure}
    \quad
    \begin{subfigure}{0.48\textwidth}
    \centering
    \includegraphics[width = 8.0cm]{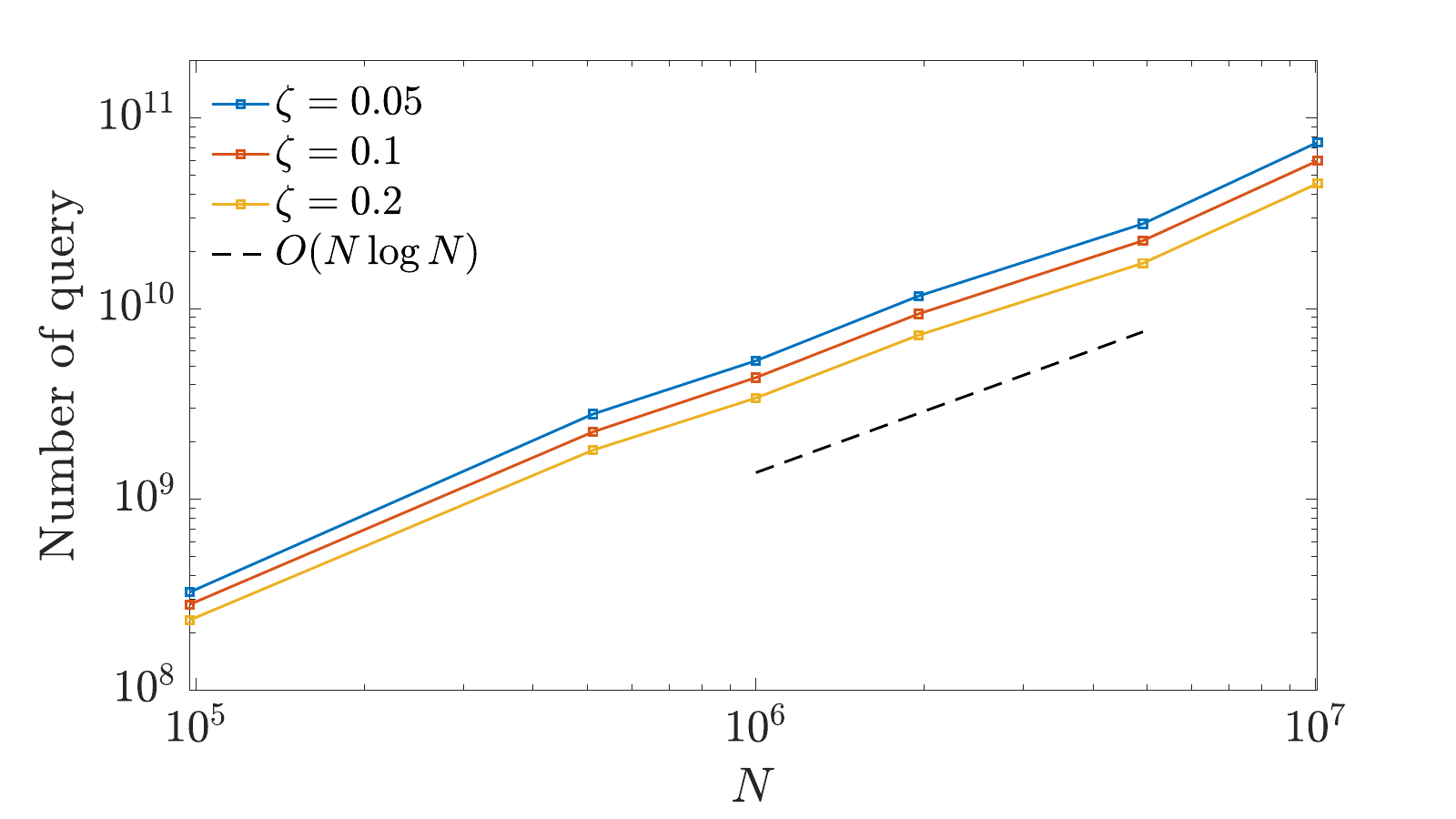}
    \caption{Fixed domain with increasing volume fraction}
    \label{sfig:complexity_zeta_fixed_domain}
    \end{subfigure}
    \caption{Number of query vs. $N$ under different $\zeta$, with $C_\text{leaf} = 50$. The black dashed line denotes the scaling of $O(N\log N)$. } 
    \label{fig:complexity_zeta}
\end{figure}

\begin{figure}[htbp]
    \centering
    \begin{subfigure}{0.48\textwidth}
    \centering
    \includegraphics[width = 8.0cm]{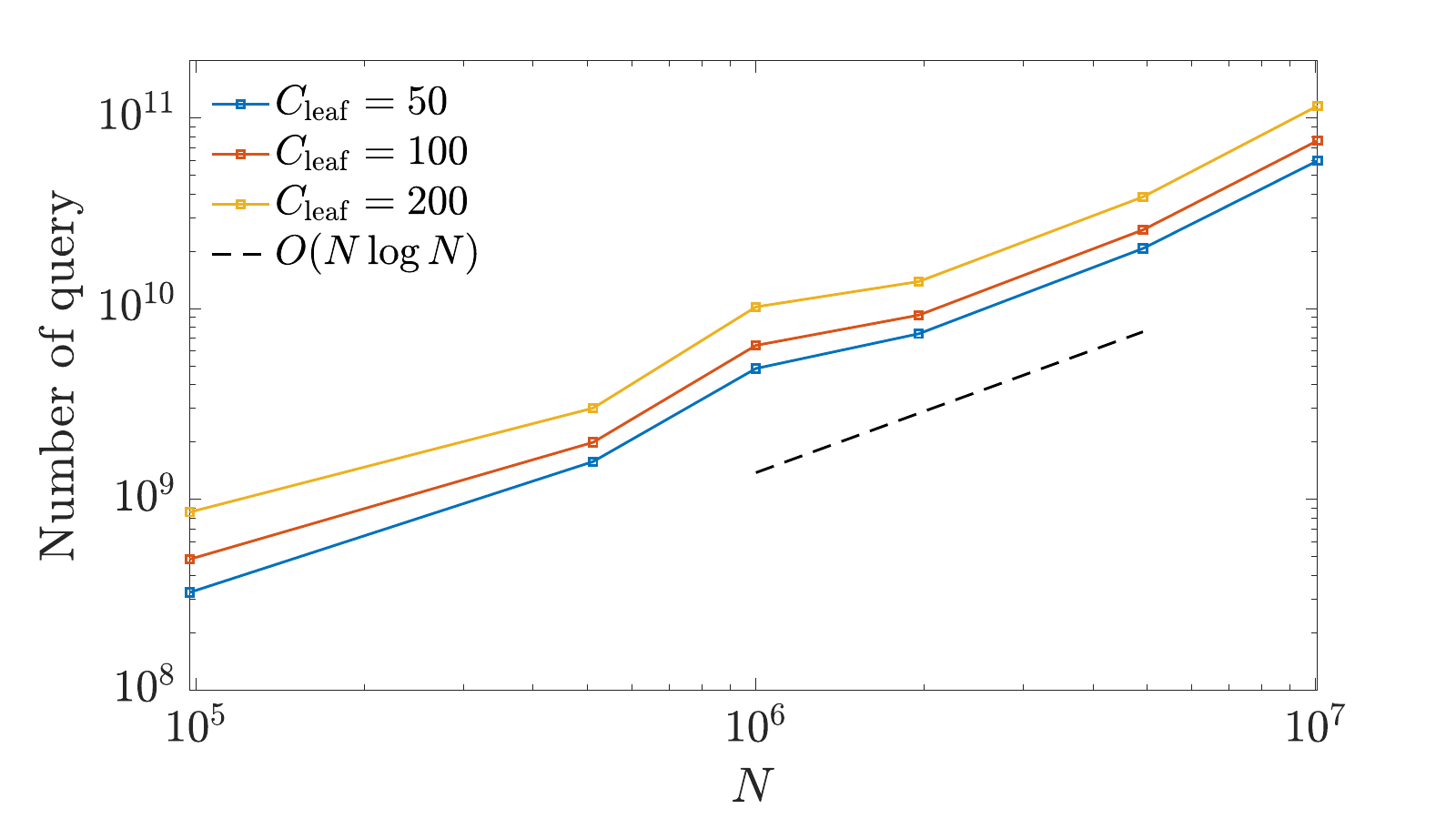}
    \caption{Fixed volume fraction with increasing domain size}
    \label{sfig:complexity_Cleaf_fixed_density}
    \end{subfigure}
    \quad
    \begin{subfigure}{0.48\textwidth}
    \centering
    \includegraphics[width = 8.0cm]{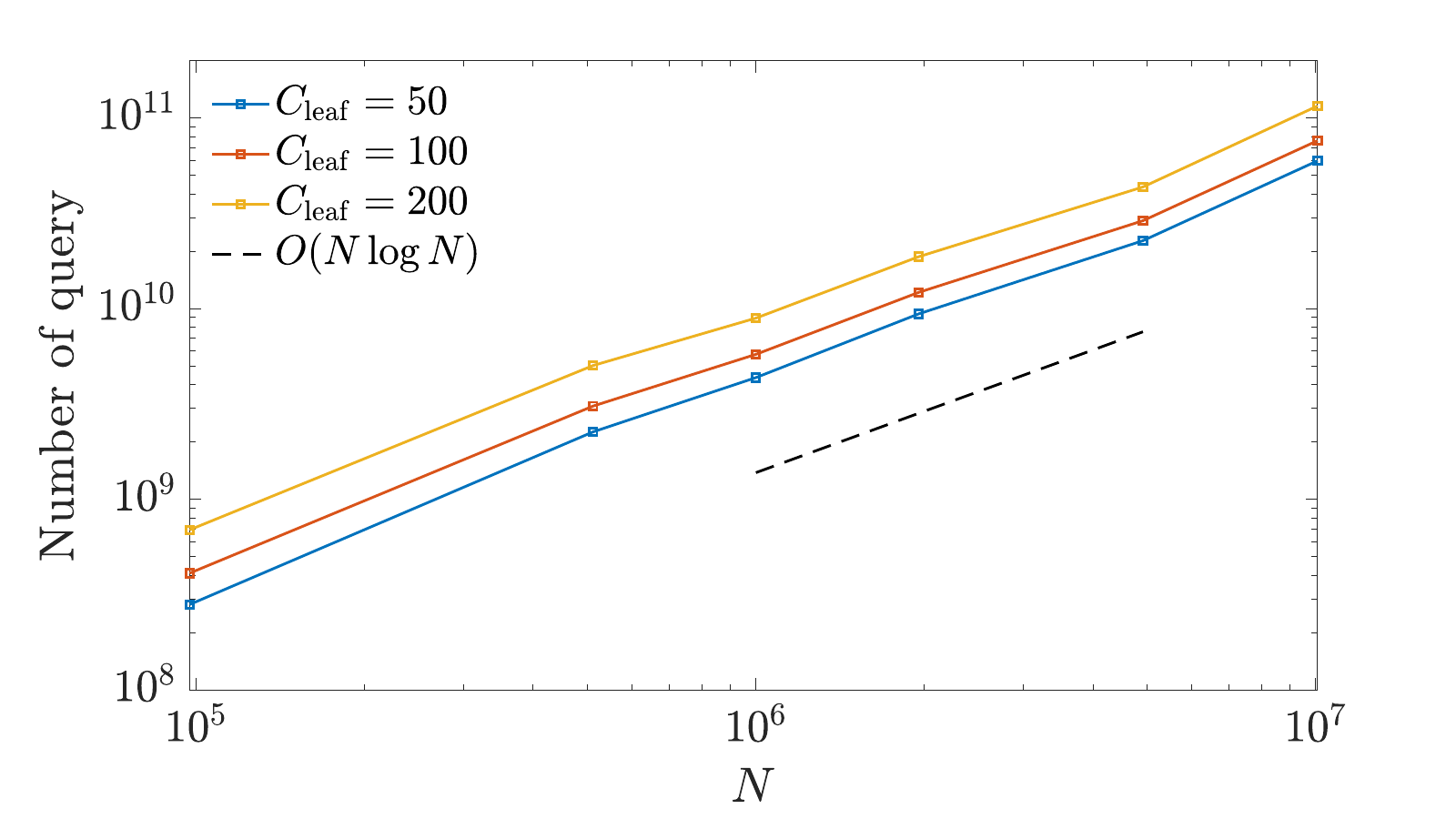}
    \caption{Fixed domain with increasing volume fraction}
    \label{sfig:complexity_Cleaf_fixed_domain}
    \end{subfigure}
    \caption{Number of query vs. $N$ under different $C_\text{leaf}$, with $\zeta = 0.1$. The black dashed line denotes the scaling of $O(N\log N)$.} 
    \label{fig:complexity_Cleaf}
\end{figure}

Finally, to determine the actual computational cost and assess the scalability of our proposed method and parallel implementation, we examined wall time as a function of particle count and GPU number. All tests were conducted on a single AWS EC2 G6E instance equipped with up to 8 NVIDIA I40 GPUs, each with 48 GB of graphics memory. By scaling the domain size while maintaining fixed particle configurations and a constant volume fraction, we obtained the performance results presented in Figure~\ref{fig:wall_time}. Specifically, 
Figure~\ref{sfig:time_vs_N} demonstrates that the wall-time closely follows the theoretical $O(N\log N)$ scaling, confirming that the computational advantages of integrating $\mathcal{H}$-matrix are preserved in our code implementation. The cluster tree branches are partitioned to achieve equitable workload distribution across a scalable number of GPUs. To assess the parallel efficiency, we further conducted the strong scalability tests using two large-scale systems containing 5 million and 10 million particles, respectively, with results shown in Figure~\ref{sfig:time_vs_GPU}. For the system as large as containing 10 million particles, the $\mathcal{H}$-HIGNN model exhibits an impressive strong scaling behavior and computational efficiency, with wall times of 1446.1 sec, 728.0 sec, 364.7 sec, and 188.6 sec for 1, 2, 4, and 8 GPUs, respectively. These results demonstrate a nearly ideal strong scaling efficiency. This can be attributed to the fact that no communication between GPUs are required for neural network inference and low-rank approximation via ACA. This parallel efficiency and scalability performance, combined with its accuracy, suggest that our proposed $\mathcal{H}$-HIGNN is well-suited for practical applications requiring the simulation of large-scale particulate systems, with the potential to handle even larger systems through distributed computing resources.
\begin{figure}[htbp]
    \centering
    \begin{subfigure}{0.48\textwidth}
    \centering
    \includegraphics[width = 8.0cm]{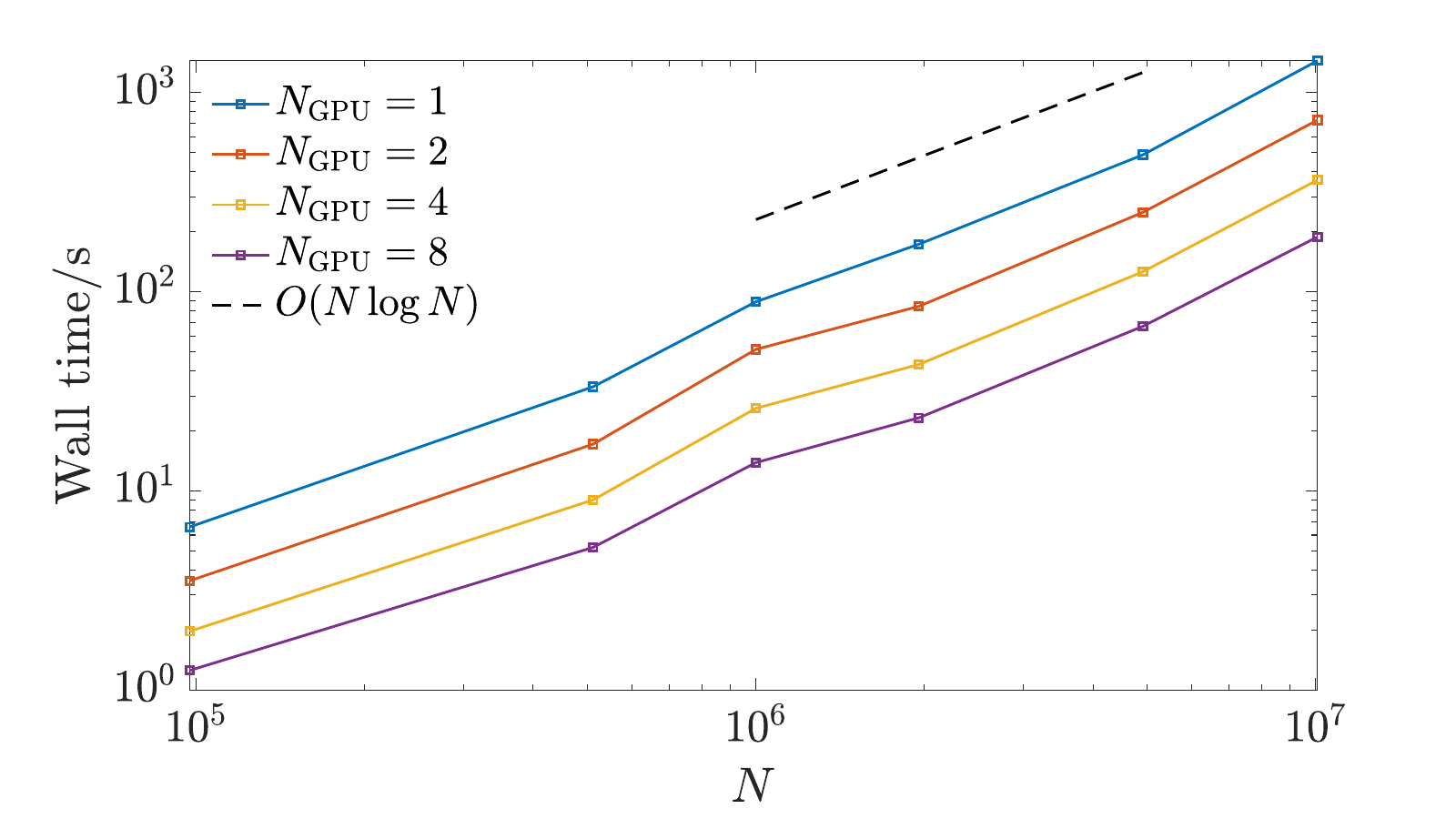}
    \caption{Wall time vs $N$ on different numbers of GPUs. The black dashed line denotes the scaling of $O(N\log N)$.}
    \label{sfig:time_vs_N}
    \end{subfigure}
    \quad
    \begin{subfigure}{0.48\textwidth}
    \centering
    \includegraphics[width = 8.0cm]{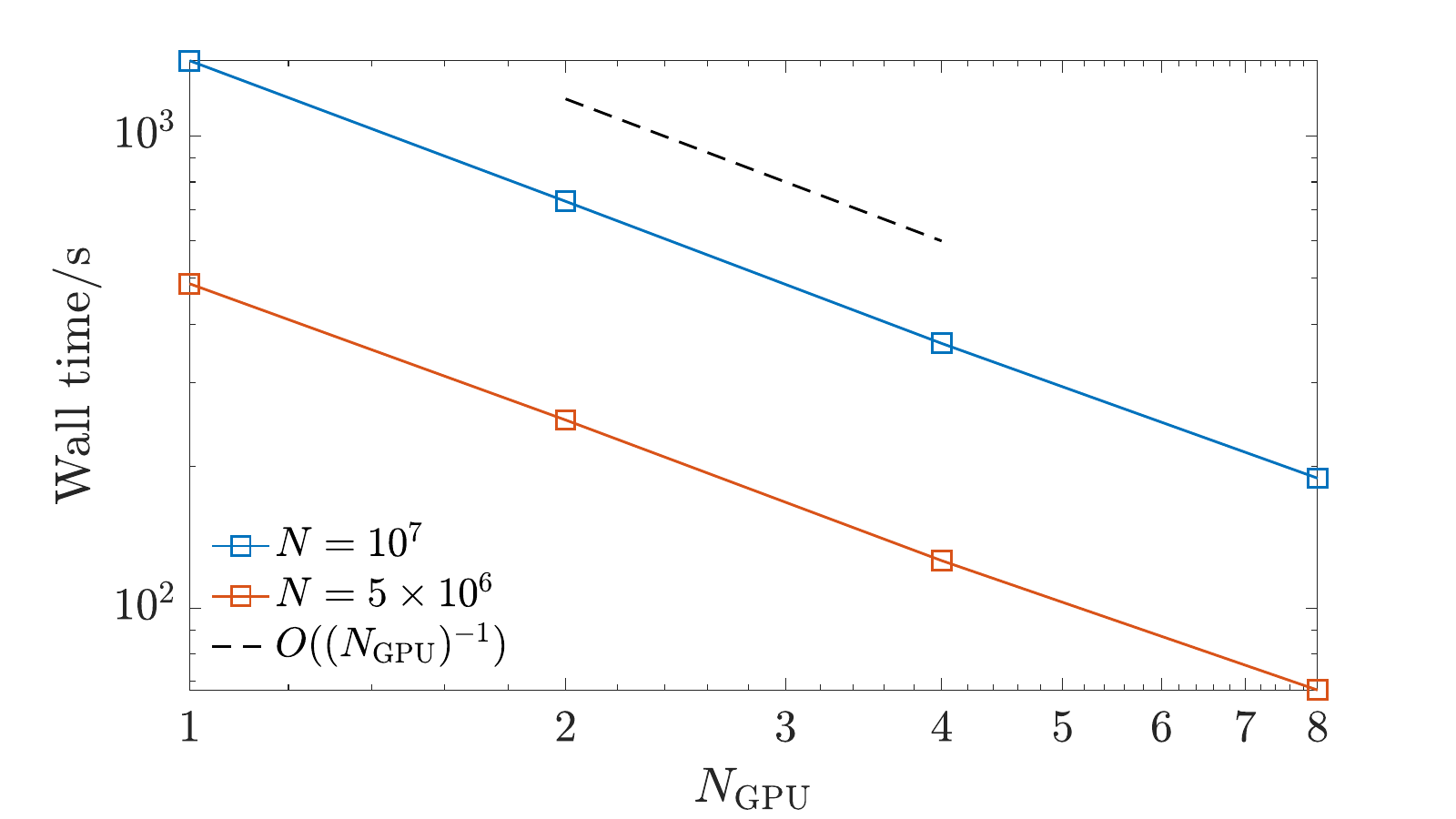}
    \caption{Wall time vs $N_\text{GPU}$ for large-scale problems. The black dashed line denotes the strong scalability with respect to $N_\text{GPU}$.}
    \label{sfig:time_vs_GPU}
    \end{subfigure}
    \caption{Scalability test of wall time with respect to $N$ (number of particles), ranging from $10^5$ to $10^7$, and $N_\text{GPU}$ (number of GPUs), ranging from 1 to 8 on a single node. Here, $\zeta = 0.1 $ and $C_\text{leaf}= 50 $}. 
    \label{fig:wall_time}
\end{figure}

It is also worth mentioning that $\mathcal{H}-$HIGNN can effectively leverage the GPU's high memory bandwidth, even when the problem is scaled up to $10^7$ particles and solved on a single GPU. This is attributed to its strategy of partitioning the mobility matrix into small blocks and performing matrix-vector multiplications sequentially on each block. As a result, at any given time step, a specific block of the mobility matrix only needs to reside in GPU memory while its computations are performed, not during operations on other blocks. This short memory lifetime for each block significantly reduces the overall GPU memory required, enabling large-scale problem-solving. In contrast, a recent GPU implementation of FSD \cite{torre2025python} requires assembling and storing the entire mobility matrix in GPU memory beforehand, as it deals with a full linear system at each time step. Thus, when the matrix size exceeds GPU memory limits, the high bandwidth cannot be exploited, which in turn restricts FSD simulations on a single GPU to the order of $10^4$ particles \cite{torre2025python}.

\subsection{Application in large-scale suspension simulation}

In this section, we present the results of applying our scalable $\mathcal{H}$-HIGNN to simulating suspensions of particles and flexible filaments, demonstrating its efficacy in addressing practically relevant large-scale problems. Note that the trained HIGNN is transferable to arbitrary numbers and configurations of particles, hence it can be directly applied without retraining. During the simulation, at each time step, $\mathcal{H}$-HIGNN was applied to predicting all particles' velocities, and the particles' positions were then updated using the explicit Euler scheme. By such, we resolved each particle's dynamics while predicting the time evolution of suspensions' macroscopic behaviors. As stated previously in \S\ref{subsec:Training_inference_HIGNN}, all numerical values are non-dimensional, with particle radius and gravitational force normalized to 1.

\subsubsection{Suspension drops}
In the first case, we simulated the interaction of two large suspension drops while sedimenting subject to gravity. Composed of a cloud of particles, each suspension drop initially formed a sphere with a radius of 175, with the volume fraction of particles as 10.0\%. 
We constructed the initial spherical suspension drop by  first generating a primitive cubic lattice structure, then discarding the particles lying outside the desired spherical domain, and finally slightly perturbing the position of each particle randomly to emulate a real suspension drop more accurately. Each suspension drop is composed of 523,305 particles, and hence the entire system contains totally 1,046,610 (more than 1 million) particles. The gravitational force acting on each particle is given by $\mathbf{F}=[0, 0, -1]$, directed along $-z$ direction. For this simulation, the time step was set to $\Delta t=0.01$, and the $\mathcal{H}$-matrix parameters were set to $\zeta = 0.1$ and $C_\text{leaf}=50$.

As shown in Figure \ref{fig:two_drops}, initially, the two suspension drops are perfectly aligned vertically without any horizontal offset. While sedimenting, due to HIs, the trailing drop is highly elongated before it catches up with the leading drop and pokes through. Later, the interface between the two drops becomes so convoluted that one cannot distinguish one drop from the other, i.e., the two drops coalesce into a bigger drop. In Figure \ref{fig:two_drops}, we present selected snapshots of the coalescence process of the two suspension drops. These results, predicted by $\mathcal{H}$-HIGNN, agree closely with the previous experimental findings reported in \cite{machu2001drop, Schaflinger1999ExpTwoDrops} and our prior HIGNN simulations for smaller-scale suspension drops \cite{ma2024shape}. The simulation was performed on a workstation with 4 NVIDIA RTX A4500 GPUs, each with 20 GB of graphics memory. It spent 28.8 sec to predict the velocities of all particles at each time step, and the entire simulation ran for 50 time steps and cost 24 min.
\begin{figure}[htbp]
    \centering
    \includegraphics[width = 0.675\textwidth]{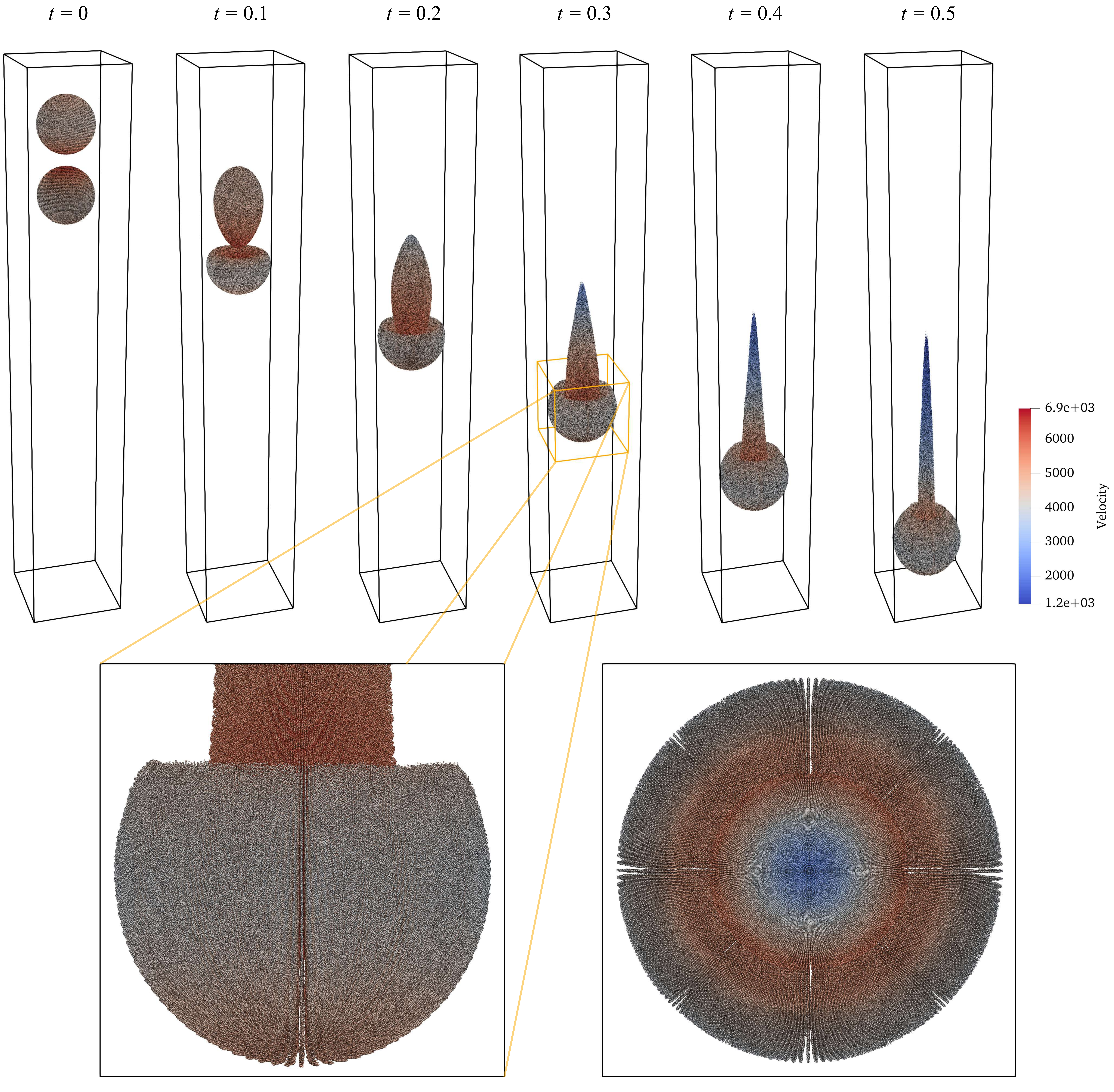}
    \caption{Snapshots of two large suspension drops, each consisting of 523,305 particles, sedimenting in an unbounded domain, predicted by $\mathcal{H}$-HIGNN. The sequence illustrates the coalescence process over time. As sedimentation progresses, the trailing drop elongates before coalescing with the leading drop. The bottom insets show a zoomed-in view and a top-down perspective at $t = 0.3$. The boxes are drawn as visual guides, and the particles' color indicates their instantaneous velocity magnitudes.}
    \label{fig:two_drops}
\end{figure}

\subsubsection{Flexible filaments}
In the second case, we increased the simulation's complexity by including flexible filaments into the particle suspension. Filaments were represented as chains of beads (particles) interconnected by elastic bonding and bending forces \cite{pan2008hydrodynamic,isele2015self,dunajova2023chiral,fox2024data}. A key feature of our model is its transferability across different types and magnitudes of external forces. Thus, the pre-trained HIGNN model can directly accommodate filaments without requiring retraining or fine-tuning, and we only introduce additional elastic bonding and bending forces to particles forming filament chains. All particles, including the beads in filaments, are identical and subjected to a uniform gravitational force $\mathbf{F}=[0, 0, -1]$, as in the previous section.
 
The elastic bonding (tension) force, applied between adjacent particle pairs $(i, i+1)$ along each filament chain, is given by \cite{isele2015self,dunajova2023chiral,fox2024data}: 
\begin{equation}
    \mathbf{f}^\text{tension}_{i\to i+1}=k_t(|\mathbf{r}_{i\to i+1}|-l_0)\frac{\mathbf{r}_{i\to i+1}}{|\mathbf{r}_{i\to i+1}|}\;, \quad i=1, 2, \dots, N_b-1 \;,
\end{equation}
where $\mathbf{r}_{i\to i+1}=\mathbf{X}_{i+1}-\mathbf{X}_i$; $l_0$ is the rest length and equal to $r_b$, the initial separation distance set between any two adjacent beads; and $k_t$ is the tension stiffness coefficient and was set to $k_t=20$ in this work. This bonding force is added to the associated particles $i$ and $i+1$ by: $\mathbf{F}_i\leftarrow\mathbf{F}_i+\mathbf{f}^\text{tension}_{i\to i+1}$ and $\mathbf{F}_{i+1}\leftarrow\mathbf{F}_{i+1}-\mathbf{f}^\text{tension}_{i\to i+1}$, respectively. 
The bending force acts on three consecutive beads $(i-1, i, i+1)$ along each filament chain and is derived from the COS bending potential \cite{pan2008hydrodynamic,fox2024data}:
\begin{equation}
    \mathcal{U}^\text{bend}_\text{COS}=k_b(1+\cos \theta) \;,
    \label{eq:bend_potential}
\end{equation}
where $k_b$ is the bending stiffness coefficient and was set to $k_b=50$ in this work; $\theta$ is the angle formed by the three consecutive beads. Specifically, the bending force is determined as follows. First, two unit tangent vectors are evaluated as: $\mathbf{t}_1=\frac{\mathbf{r}_{i-1\to i}}{|\mathbf{r}_{i-1\to i}|}$ and 
$\mathbf{t}_2=\frac{\mathbf{r}_{i\to i+1}}{|\mathbf{r}_{i\to i+1}|}$. The angle between the two unit tangent vectors can then be determined from: $\cos \theta_i =\mathbf{t}_1 \cdot \mathbf{t}_2$ and $\sin \theta_i = |\mathbf{t}_1 \times \mathbf{t}_2|$. Next, the torque due to bending is computed as: $\mathbf{T}^\text{bend}_i=-k_b(\mathbf{t}_1 \times \mathbf{t}_2)\sin \theta_i$. The forces induced by the bending torque on each of the three associated beads are then given by: $\mathbf{f}^\text{bend}_{i-1}=\mathbf{T}^\text{bend}_i\times \mathbf{t}_1$, $\mathbf{f}^\text{bend}_{i+1}=-\mathbf{T}^\text{bend}_i\times \mathbf{t}_2$, and $\mathbf{f}^\text{bend}_{i}=-(\mathbf{f}^\text{bend}_{i-1}+\mathbf{f}^\text{bend}_{i+1})$, respectively. Finally, these forces are added to the associated particles by: $\mathbf{F}_{i-1}\leftarrow\mathbf{F}_{i-1}+\mathbf{f}^\text{bend}_{i-1}$, $\mathbf{F}_i\leftarrow\mathbf{F}_i+\mathbf{f}^\text{bend}_{i}$ and $\mathbf{F}_{i+1}\leftarrow\mathbf{F}_{i+1}+\mathbf{f}^\text{bend}_{i+1}$, respectively.  

Initially, a spherical suspension of a mixture of particles and filaments was constructed as follows. The sphere's radius was set to $R_\text{out} = 4 (10)^{\frac13}(N_b - 1)r_b$, where $N_b$ denotes the number of beads on each filament. Thus, the filament length $L_f$ can be estimated from $L_f = (N_b - 1)r_b+2a$ with $a=1$ the radius of a bead (particle). In this work, each filament comprises $N_b=21$ beads with $r_b = 3$, yielding $L_f=62$. Filaments were randomly oriented and positioned within an inner sphere of radius $R_\text{in} = {(\frac12)}^{\frac13} R_\text{out}$. Individual particles were randomly distributed within the spherical shell between radii $R_\text{in}$ and $R_\text{out}$. To prevent overlap, the initial minimum distance between individual particles and/or filaments was set to 2.1. The simulation involved 5120 filaments and 107,520 individual particles, resulting in an equal number of particles in filaments and individual particles, for a total of 215,040 particles simulated. The initial configuration of this suspension containing flexible filaments and individual particles is depicted in Figure \ref{fig:sus_filament_t=1}.  
\begin{figure}[htbp]
\centering
\includegraphics[width = 1.0\linewidth ]{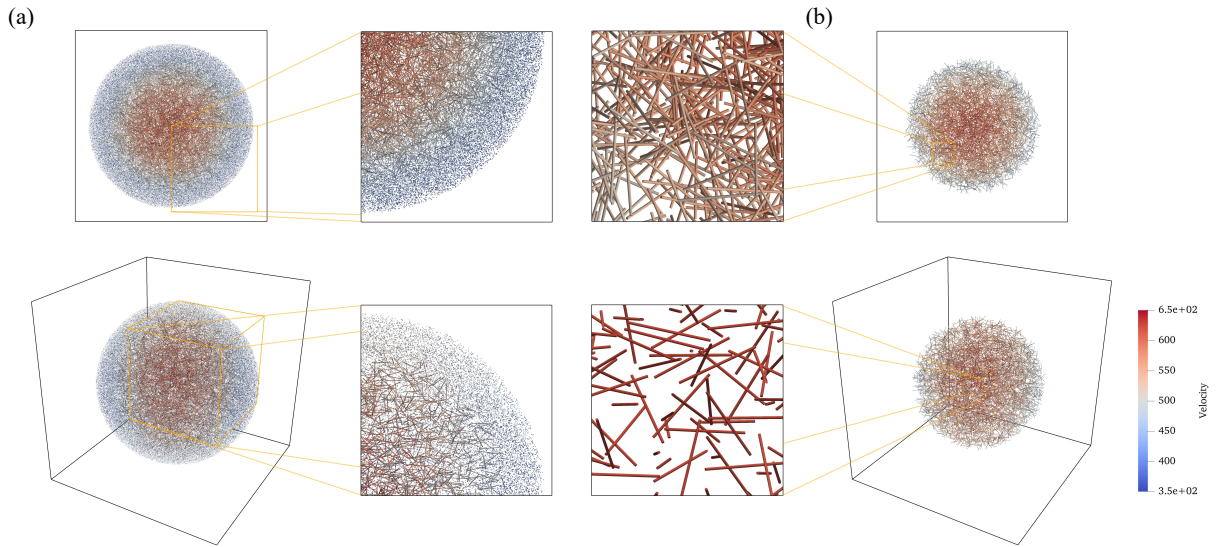}
\caption{Initial configuration of a spherical suspension containing 5,120 flexible filaments and 107,520 individual particles subjected to a uniform gravitational force. (a) Full suspension including both filaments and particles. (b) Filaments isolated for clarity, highlighting their random orientations and centralized distribution. Both panels include 3D (bottom row) and top-down (top row) views. Filaments are initially concentrated near the center of the cloud and exhibit higher downward velocities. In contrast, particles located in the outer layers experience lower downward velocities. The boxes are drawn as visual guides, and the color scale represents instantaneous velocity magnitudes.}
\label{fig:sus_filament_t=1}
\end{figure}

Same as in the prior section, the simulation was conducted on a workstation with 4 NVIDIA RTX A4500 GPUs, each with 20 GB of graphics memory. For this simulation, the time step was set to $\Delta t = 0.002$, and the $\mathcal{H}$-matrix parameters were set to $\zeta = 0.02$ and $C_\text{leaf}=50$. The wall time for each time step was about 8.3 sec. The simulation was run up to the final time $T = 20$. 

While all particles, including those within filaments, are subject to identical gravitational forces, HIs result in spatially varying downward velocities. As filaments are initially located in the cloud's center, their higher downward velocities drive them towards the cloud's front boundary. The particles initially in the cloud's outer layers, including the front boundary, experience lower downward velocities and are swept towards the rear. As a result, filaments are observed to move from the center to the front boundary, then to the rear, as illustrated in Figures \ref{fig:sus_filament_t=2} and \ref{fig:sus_filament_t=3}. In this context, ``front" and ``rear" are oriented according to the sedimentation direction.
\begin{figure}[htbp]
\centering
\includegraphics[width = 1.0\linewidth ]{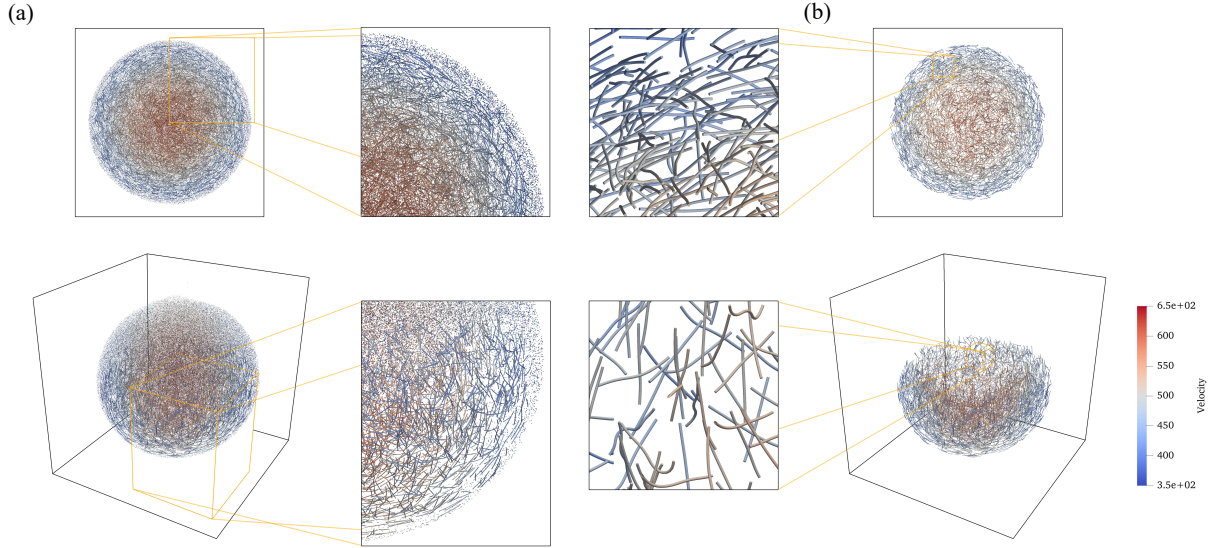}
\caption{The suspension in Figure \ref{fig:sus_filament_t=1} sedimenting in an unbounded domain: Snapshot at $t = 6$ predicted by $\mathcal{H}$-HIGNN. (a) Full suspension including both filaments and particles. (b)  Filaments isolated for clarity, highlighting their deformations and spatial distribution. Both panels include 3D (bottom row) and top-down (top row) views. The filaments, initially centralized, advance toward the front boundary of the cloud due to their higher downward velocities, while individual particles from the cloud's outer layers experience lower downward velocities and are swept towards the rear. The boxes are drawn as visual guides, and the color scale represents instantaneous velocity magnitudes.}
\label{fig:sus_filament_t=2}
\end{figure}
\begin{figure}[htbp]
\centering
\includegraphics[width = 1.0\linewidth ]{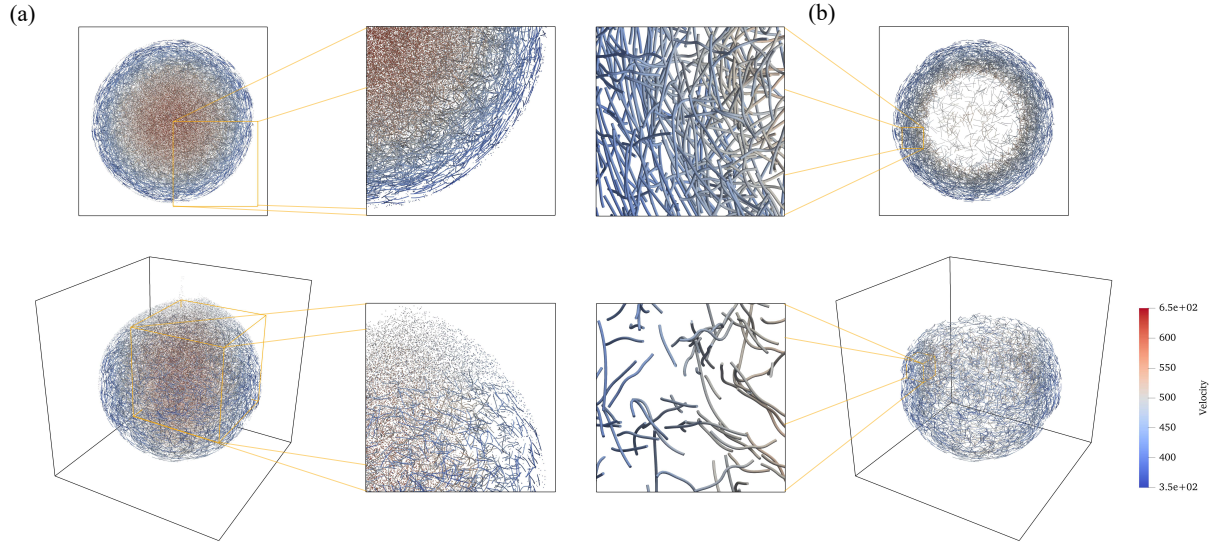}
\caption{The suspension in Figure \ref{fig:sus_filament_t=1} sedimenting in an unbounded domain: Snapshot at $t = 11$ predicted by $\mathcal{H}$-HIGNN. (a) Full suspension including both filaments and particles. (b) Filaments isolated for clarity, highlighting their deformations and spatial distribution. Both panels include 3D (bottom row) and top-down (top row) views. The filaments that previously reached the front boundary of the cloud move toward the rear due to their reduced downward velocities, forming an outer shell. In contrast, individual particles concentrate near the center of the cloud and experience higher downward velocities. The boxes are drawn as visual guides, and the color scale represents instantaneous velocity magnitudes.}
\label{fig:sus_filament_t=3}
\end{figure}
\begin{figure}[htbp]
\centering
\includegraphics[width = 1.0\linewidth ]{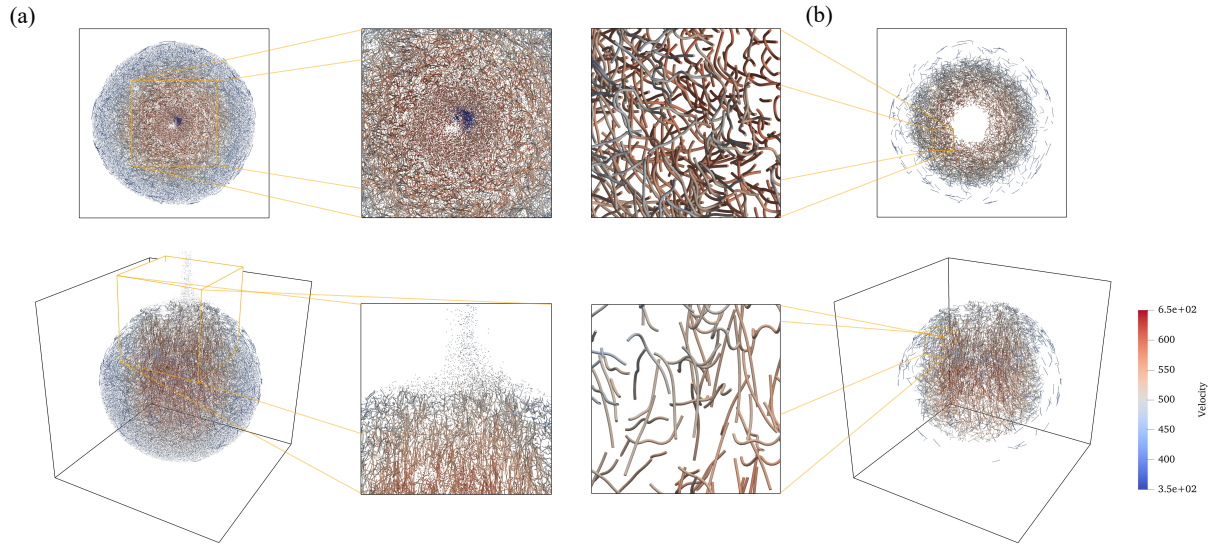}
\caption{The suspension in Figure \ref{fig:sus_filament_t=1} sedimenting in an unbounded domain: Snapshot at $t = 20$ predicted by $\mathcal{H}$-HIGNN. (a) Full suspension including both filaments and particles. (b)  Filaments isolated for clarity, highlighting their deformations and spatial distribution. Both panels include 3D (bottom row) and top-down (top row) views. The filaments that previously moved to the outer layers of the cloud convolute back toward the center and experience higher downward velocities, while individual particles near the center are pushed to the outer layers and exhibit reduced downward velocities. Filaments are more effectively retained within the circulatory region created by the Hill’s vortex near the rear of the cloud, while particles outside this region are carried away by the background flow, forming an axial tail. The boxes are drawn as visual guides, and the color scale represents instantaneous velocity magnitudes.}
\label{fig:sus_filament_t=4}
\end{figure}

Near the cloud's rear, as Hill's vortex forms an envelope of enclosed streamlines \cite{NITSCHE_BATCHELOR_1997,Ayeni2020ParticleCloud}, particles and filaments within this envelope circulate towards the center, establishing a circulatory motion within the sedimenting cloud. Particles outside this envelope, however, are caught in the background fluid's backward sweep, resulting in particle leakage from the cloud's rear as an axial tail \cite{NITSCHE_BATCHELOR_1997,Ayeni2020ParticleCloud}. Compared to individual particles, filaments are more readily captured by the enclosed streamlines, thus maintaining a circulatory motion and moving from the rear towards the center, as seen in Figure \ref{fig:sus_filament_t=4}. The vertical tail shed behind the cloud primarily consists of individual particles. 

All these critical phenomena were correctly captured in our predictions by $\mathcal{H}$-HIGNN, as shown in Figures \ref{fig:sus_filament_t=2}-\ref{fig:sus_filament_t=4}. Further, due to their flexibility and HIs with the surrounding fluid and other immersed objects, filaments exhibited various deformed shapes during this dynamical process. These shapes include V-, U-, C-, S-, and W-shaped modes, as reported in the literature \cite{du2019dynamics,schoeller2021methods,bonacci2023dynamics} and illustrated in Figures \ref{fig:sus_filament_t=2}-\ref{fig:sus_filament_t=4}.

\section{Conclusion}\label{sec:conclusion}
We have presented a fast and scalable computational framework, $\mathcal{H}$-HIGNN, for simulating large-scale particulate suspensions. It inherits all advantages of HIGNN, including the capability of resolving short- and long-range, many-body HIs, computational efficiency, explainability, and transferability. By integrating $\mathcal{H}$-matrix techniques, it further improves the scalability. In particular, the incorporation of $\mathcal{H}$-matrix reduces the computational complexity of matrix-vector multiplication in Eq. \eqref{Eq:theo_basis_mat} to quasi-linear, thereby yielding a quasi-linear scaling for the prediction cost and enhancing the overall scalability. The key ingredients of $\mathcal{H}$-HIGNN, alongside GNNs, include: 1) cluster-tree block partitioning using geometric information embedded in the graph; 2) cluster distance-to-size ratios for admissibility determination; and 3) ACA for low-rank approximations of admissible blocks. 

We note that the particle–mesh Ewald (PME) method \cite{darden1993particle,yeh1999ewald,brodka2004ewald} and the fast multipole method (FMM) \cite{greengard1987fast,ying2004kernel,yokota2015fast} may potentially be adapted to address our problem. However, we adopted the $\mathcal{H}$-matrix approach based on our expertise. Moreover, $\mathcal{H}$-matrix tends to offer better parallel efficiency \cite{brunner2010comparison}, which is critical for the large-scale systems considered in this work.

We have systematically examined the accuracy of $\mathcal{H}$-HIGNN against HIGNN in the present work, given HIGNN's thorough validation in our prior work \cite{Ma2022HIGNN,ma2024shape}. We have also validated $\mathcal{H}$-HIGNN's $O(N\log N)$ computational complexity and scalability through a comprehensive evaluation. By optimizing GPU latency, workload distribution, and inter-GPU communication, and leveraging the Kokkos programming model \cite{9485033}, our GPU implementation achieves near-theoretical $O(N\log N)$ wall-time scaling and demonstrates near-ideal strong scalability. This work, to our best knowledge, represents the first demonstration of linearly scalable computation for suspensions as large as comprising $10^7$ particles, utilizing minimal computing resources like a single GPU.

To showcase $\mathcal{H}$-HIGNN's practical application, we also have presented the results of applying our scalable $\mathcal{H}$-HIGNN to efficiently simulating practically relevant large-scale suspensions. Due to HIGNN's inherent transferability across arbitrary numbers and configurations of particles and external forces, it can be directly applied without retraining or any fine-tuning. Simply by adding elastic bonding and bending forces on particles, we have also extended $\mathcal{H}$-HIGNN to simulating large-scale suspensions of more than 5000 flexible filaments. This extension of our framework could unlock its potential for broader applications to more complex physical systems where filaments are constituent elements, such as networks and membranes. Moreover, by incorporating self-propulsion forces or external fields, our scalable $\mathcal{H}$-HIGNN framework can efficiently simulate active matter and self-propelled microswimmers, potentially at scales previously inaccessible.

\section*{CRediT authorship contribution statement}
\textbf{Zhan Ma}: Writing – original draft, Conceptualization, Methodology, Software, Investigation, Validation, Formal analysis, Data curation. \textbf{Zisheng Ye}: Writing – original draft, Methodology, Software, Investigation, Validation, Formal analysis, Data curation.
\textbf{Ebrahim Safdarian}: Writing – original draft, Investigation, Visualization, Formal analysis. 
\textbf{Wenxiao Pan}: Writing – original draft, Writing – review \& editing, Supervision, Conceptualization, Methodology, Investigation, Funding acquisition, Resources, Project administration.

\section*{Declaration of competing interest}
The authors declare that they have no known competing financial interests or personal relationships that could have appeared to influence the work reported in this paper.

\section*{Acknowledgments}
We gratefully acknowledge the funding support for this work provided by Army Research Office Grant \\
No. W911NF2310256.

\section*{Data Availability}
The codes and examples that support the findings of this study are available in the link: \url{https://github.com/Pan-Group-UW-Madison/hignn}.

\bibliographystyle{elsarticle-num}
\bibliography{ref}
\biboptions{sort&compress}
\end{document}